**Tuning of Magnetic and Electrical Properties in Complex oxide Thin Films Deposited By Pulsed Laser Deposition**

A Thesis
Presented to
The Academic Faculty

by

Arun Singh Chouhan

In Partial Fulfillment of the Requirement for the Degree of M.Tech.
(Nano Science & Technology)

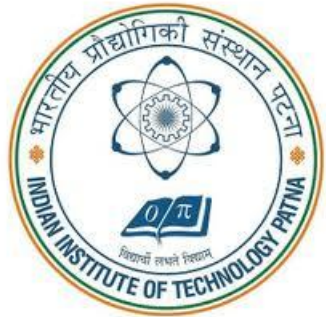

**Indian Institute of Technology Patna
India**

April 2015

**Copyright © Arun Singh Chouhan 2015**

***Dedicated to my beloved parents and my lovely sister***

*For their endless love and support*

***Also dedicated to my best friend Nidhi***

*Who has been a great source of motivation & inspiration*

INDIAN INSTITUTE OF TECHNOLOGY PATNA

**Evaluation Committee Approval**

of a Thesis submitted by

ARUN SINGH CHOUHAN

The thesis of **Arun Singh Chouhan** has been read, found satisfactory and approved by the following DPPC committee members.

Name : 1______________________,
Coordinator School of Engineering/Basic Sciences   Date : ________________

Name : 2 ______________________, Supervisor     Date : ________________

Name : 3 ______________________, Co-Supervisor  Date : ________________

Name : 4 ______________________, Member         Date : ________________

Name : 5 ______________________, Member         Date : ________________

Name : 6 ______________________, Member         Date : ________________

Name : 7 ______________________, Member         Date : ________________

Name : 8 ______________________, Member         Date : ________________

# ACKNOWLEDGEMENT

This M.Tech. thesis work would not be possible without the help and guidance of a number of people. It is now opportune moment for me to acknowledge my gratitude for all of them.

I am greatly indebted to my thesis supervisors Dr. Ajay D Thakur and Prof. Dr. Lambert Alff for their inspiring guidance throughout my research work. It was a memorable experience to work with and learn many things from them. I am also thankful to all faculty members at Indian Institute of Technology (IIT) Patna for their immense guidance and motivation.

I thank Dr. Vivek Malik for providing many of the theoretical materials that are used in the research project.

I am thankful to Dr. Philipp Komissinskiy for the many valuable discussions with him throughout my research period at Technical university (TU) Darmstadt, Germany.

I am thankful to Dr. Erwin Hildebrandt, Mr. Supratik Das Gupta, Mr. Vikas Shabadi, Mr. Aldin Radetinac, Mr. Arzhang Mani, Mr. Sharath Ulhas, Mr. Soumya J Roy, for their valuable support throughout my stay at TU Darmstadt, Germany.

It is a pleasure thank to staff members, Mr. Santosh Kumar, Mr. Sanjeev Kumar and Ms. Shabana Tabassum from IIT Patna and Mr. Jurgen Scherech, Ms. Gabi Haindl and Ms. Marion Bracke from TU Darmstadt, Germany.

I would like to thank research scholar Mr. Ashutosh Kumar, Mr. Pranay Ranjan, Mr. Ajay Nath and Mr. Atma Rai for their support and guidance during my stay at IIT Patna.

I would also like to thank Mess staff, Hostel Staff, security people for providing good and safe environment at IIT Patna during my M.Tech. tenure.

I would like to thank DAAD for providing me funding to carry out my research work at TU Darmstadt, Germany. I would also like to thank MHRD for proving me fellowship to support my M.Tech. studies.

Finally, I would like to thank my beloved parents, my sister and my best friend for all their love and support throughout my research work.

# CERTIFICATE

This is to certify that the thesis entitled "Tuning of Magnetic and Electrical Properties in Complex oxide Thin Films Deposited By Pulsed Laser Deposition", submitted by Arun Singh Chouhan to Indian Institute of Technology Patna, is a record of bonafide research work under our supervision and we consider it worthy of consideration for the degree of Master of Technology of this Institute. This work or a part has not been submitted to any university/institution for the award of degree/diploma. The thesis is free from plagiarized material.

\_\_\_\_\_\_\_\_\_\_\_\_\_\_\_\_\_\_\_\_\_\_\_

\_\_\_\_\_\_\_\_\_\_\_\_\_\_\_\_\_\_\_\_\_\_\_
Supervisor

\_\_\_\_\_\_\_\_\_\_\_\_\_\_\_\_\_\_\_\_\_\_\_

\_\_\_\_\_\_\_\_\_\_\_\_\_\_\_\_\_\_\_\_\_\_\_
Co-Supervisor                                             Date: \_\_\_\_\_\_\_\_\_\_\_\_

# DECLARATION

I certify that

a. The work contained in this thesis is original and has been done by myself under the general supervision of my supervisors Dr. Ajay D Thakur and Prof. Dr. Lambert Alff.

b. The work has not been submitted to any other Institute for degree or diploma.

c. I have followed the Institute norms and guidelines and abide by the regulation as given in the Ethical Code of Conduct of the Institute.

d. Whenever I have used materials (data, theory and text) from other sources, I have given due credit to them by citing them in the text of the thesis and giving their details in the reference section.

e. The thesis document has been thoroughly checked to exclude plagiarism.

<div style="text-align: right;">
Signature of the Student  
Roll No:
</div>

# TABLE OF CONTENT

# LIST OF FIGURES

# LIST OF TABLE

**Table 3.9.2.1** : Magnetic moment for LMO thin films grown in argon with different fluence at same pressure.

# LIST OF ABBREVIATIONS

| | |
|---|---|
| **CMR** | Colossal Magneto Resistance |
| **PLD** | Pulsed Laser Deposition |
| **SE** | Super Exchange |
| **DE** | Double Exchange |
| **EDXA** | Energy Dispersive X-ray Analysis |
| **SQUID** | Super Conducting Quantum Interference Device |
| **XRD** | X-Ray Diffraction |
| **XRR** | X-Ray Reflectivity |
| **AFM** | Anti-Ferro Magnetic |
| **FM** | Ferro Magnetic |
| **f.u.** | Formula Unit |
| **CFSE** | Crystal Field Stabilization Energy |
| **FC** | Field Cooled |
| **ZFC** | Zero Field Cooled |
| **RHEED** | Reflection High Energy Electron Diffraction |
| **eV** | Electron Volt |
| **mBar** | Milli Bar |

# ABSTRACT

Lanthanum manganite, LaMnO$_3$ (LMO) is the parent compound for a class of hole doped (e.g. La$_{1-x}$Ca$_x$MnO$_3$, La$_{1-x}$Sr$_x$MnO$_3$) and electron doped (e.g. . La$_{1-x}$Ce$_x$MnO$_3$, La$_{1-x}$Sn$_x$MnO$_3$) perovskite complex oxide materials. Strong correlation between the spin, lattice, charge and orbital degrees of freedom is a hallmark of these class of materials (popularly known as manganites). Competition and interplay of these degrees of freedom lead to a wide range of interesting electronic behavior including half-metallicity, colossal magneto-resistivity, etc [1-6]. Although a lot of experimental research has been carried out on the hole doped and the electron doped counterparts, the parent system LMO remains less investigated [7,8]. This could be attributed to the difficulty in making stoichiometric LMO (whether in bulk polycrystalline, single crystals or thin films). There exists phenomenon like double-exchange (DE) and anti-ferromagnetic super-exchange (SE) interactions which leads to ferromagnetism in LMO thin films. This project work is aimed at the fabrication and characterization of thin films of stoichiometric LMO and their structural, magnetic and electrical studies.

The LMO target were prepared by solid state synthesis using 4N (99.99% pure) La$_2$O$_3$ and Mn$_2$O$_3$ as precursors. Respective phase and stoichiometry of target were confirmed by using X-ray diffraction (XRD) and energy dispersive X-ray analysis (EDXA). Thin films were deposited on 001 SrTiO$_3$ (STO) substrates using KrF pulsed laser deposition. The parameters like laser fluence, pulse frequency, gas environment, substrate temperature, pressure has been varied to optimize LMO thin films. Epitaxy and thickness of the films were characterized using thin film X-ray diffraction (TXRD), x-ray reflectivity (XRR) and reflective high energy electron diffraction (RHEED). Magnetic measurements were carried out using SQUID (superconducting quantum interference device) magnetometry to ascertain the magnetic behavior. Electrical transport measurements were carried out using four probe technique.

After the fundamentals of perovskite , theory of thin film growth and exchange mechanism in chapter 1, followed by all the experimental techniques used throughout the project work in chapter 2. Results are discussed in chapter 3 with conclusion and outlook are discussed in chapter 4.

# CHAPTER 1
# INTRODUCTION

**1.1. Motivation**

Perovskite oxides are unique materials having many interesting physical and chemical properties. These perovskite structures are having complex phase diagram due to various types of interactions between charge and spin in addition to various other parameters. One class of perovskite structure is rare-earth manganites ($RMnO_3$). These compounds and their chemically doped derivatives are most studied perovskite compounds due to the presence of many unique properties like colossal magneto resistance, magneto electric coupling with many temperature and pressure dependent chemical properties [1-6]. They also posses unique optical properties and can be used as optical filters. Since strongly correlated electrons in RMnO generate a large variety of competing ground states in bulk, incorporating them in heterostructures provides with additional opportunities for creating novel phenomena at interfaces. Therefore, synthesizing them in thin film form offers additional degrees of freedom. Thin films of such materials can be deposited by using many deposition techniques, one of them is pulsed laser deposition(PLD). Properties of such materials can be varied and modified by manipulating strain in the material. This can be done by depositing thin films on different substrates and using different growth conditions in case of PLD process. Same material can possesses different properties when grown with different deposition conditions, which can be used in various applications.

These compounds are having huge potential as mentioned above, but understanding of properties of $LaMnO_3$ (LMO) thin films and their heterostructures has been challenged due to the difficulty in achieving bulk like electrical and magnetic properties i.e, anti-ferromagnetic (AFM) and insulating. Most LMO thin films, regardless of the growth method, exhibit ferromagnetic (FM) insulating behaviors [10-13]. Therefore heterostructures having LMO as part are not that easily reproducible. There are many problems which are still unanswered and one of them is to reduce the magnetization of LMO thin films.

Thin film growth of perovskites ($ABO_3$) by using pulsed laser deposition (PLD) requires a simultaneous precise control of many growth parameters. Parameters like background gas, laser fluence, gas used, pulse frequency, temperature, have effect on the stoichiometry and thereby also the functionality of compounds such as $LaMnO_3$ (LMO) where the *B*-site cation is having multiple stable oxidation states. Although stoichiometric bulk LMO is an *A*-type anti-ferromagnetic insulator, several independent studies have confirmed a ferromagnetic order in thin films owing to an off-stoichiometry driven double exchange interaction between the mixed $Mn^{3+}$ and $Mn^{4+}$ oxidative states [10]. In spite of repeated efforts, perfectly stoichiometric films with magnetic saturation below ~ 0.5 $\mu_B$/f.u. has been difficult to produce without any annealing treatment.

There might be many possible reasons for this induced ferromagnetism order and its very interesting to figure out the reasons and dependence of ferromagnetic order on those. This project work is concerned with reduction of ferromagnetic order by exploring different parameters during the deposition of thin films.

**1.2. Perovskite Manganite ($LaMnO_3$)**

A perovskite structure can be any material be same type of crystal structure as calcium titanium oxide ($CaTiO_3$) [14]. The general chemical formula for perovskite compounds is $ABO_3$, where 'A' and 'B' are cations of different sizes with 'O' as anion bonds to both 'A' and 'B'. In general, A atoms are bigger than B atoms. In case of $LaMnO_3$, La is on 'A' site, Mn on 'B' site with O atoms as face centered [15].

Stoichiometric LMO is an A-type anti-ferromagnetic insulating material having $T_N$ = 140K. LMO is insulating at all temperatures and having optical bandgap of ~1.7 eV. Thin films of LMO has many applications like magnetic field sensors, field effect devices, importance in heterostructures of different magnetic material. LMO has a very rich phase diagram due to its oxygen non-stoichiometry.

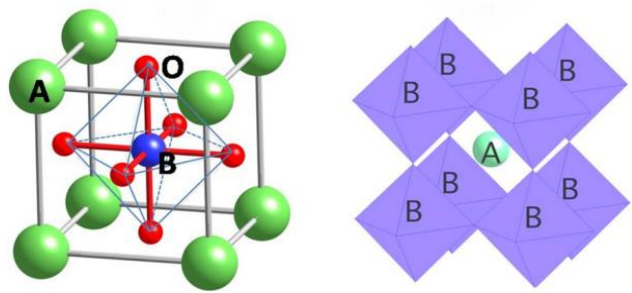

Figure 1.2.1. Structure of $ABO_3$ perovskite with origin centred at B-site cation and A-site cation

### 1.2.1. Crystal Field Theory

As explained by crystal field theory, degenerated d orbitals of transition metal splits into $t_{2g}$ and $e_g$ [16] as shown in figure 1.2.2. A splitting of energy levels occurs because of the orientation of the d orbital wave functions which will increase an electron's energy when the orbital is located in a region where electron density is high and lower it when the reverse is true. The $d_{xy}$, $d_{yz}$, $d_{xz}$, $d_{x^2-y^2}$ and $d_{z^2}$ orbitals in crystals split up as depending on cations coordination. The crystal field stabilization energy (CFSE) is a measure of stability of transition metal ion place in the crystal field generated by set of ligands. When d-orbitals splits in ligand field, some of become higher in energy with remaining lower in energy. As a result of this, if any electron occupying these orbitals, the metal ion is more stable in the ligand field relative to barycenter. The $e_g$ orbitals in case of octahedral are higher in energy than the barycenter, so putting electron in this reduces the CFSE.

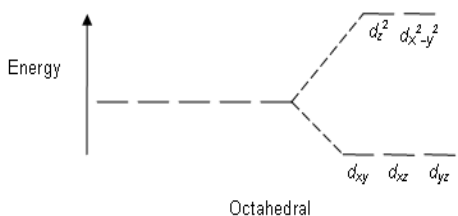

Figure 1.2.1.1. Crystal field splitting in octahedral symmetry

### 1.2.2. Exchange Mechanism

The exchange interaction is a quantum mechanical effect between identical particles. In general, these exchange mechanism defines the magnetic properties in a material. Based on interaction, exchange mechanism can be divided into two major categories, 'Direct' and 'Indirect' exchange mechanism.

**Direct Exchange Interaction :-** Direct exchange interaction takes place between moments which are close enough to make their wave functions overlap. This interaction is very strong but short range and decreases rapidly with distance.

**Indirect exchange interaction** :- It consists of three main interactions namely, RKKY, Super exchange and double exchange.

**RKKY Interaction :-** RKKY stands for Ruderman-Kittel-Kasuya-Yosida. This explains coupling of d or f shell electrons through the conduction electrons.

**Super Exchange Interaction :-** Super exchange interaction describes the interaction between moments of ions which are too far for a direct exchange interaction to happen but interact through a non magnetic ion. It was proposed by Hendrik Kramers in 1934 when he noticed this in crystals like MnO. A model for superexchange in MnO is shown in figure 1.3.1.

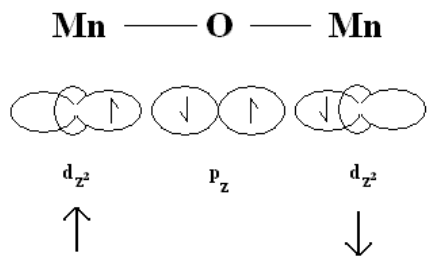

Figure 1.2.2.1. Superexchange in MnO

**Double Exchange Interaction :-** The double exchange mechanism is a type of a exchange interaction which arises betweens ions in different ionic states. Clarence Zener has introduced this term. This interaction takes place due to the hopping of electrons between ions and during the hopping sign of spin is preserved that is why interaction leads to ferromagnetism.

**1.3. Theory of thin film growth**

Based on different atomic and microscopic states of the order of growing layer, solid thin films can be categorized into four types.

**Amorphous thin films:-** Atomic order which is of very short range i.e, no ordering of atoms.

**Polycrystalline thin films**:- Atomic ordering within the grains but no ordering between the grains.

**Textured thin films**:- Atomic ordering within the grains with oriented grains along a particular direction.

**Epitaxial thin films**:- Long range atomic order in grains with orientation relation between grains and the respective substrate.

However, some special cases and mixture of these types are also possible. There are many benefits of highest ordering and epitaxial growth. It is also possible to study effect on crystal quality of a material due to the compositions of a certain constituent material. In addition, many materials can also be grown with single crystal quality irrespective of the unfavourable thermodynamics. New functionalities can also be induced by growing artificial materials through growth of layered structures of different materials. Few terms related to thin film growth and very important also are described below.

Epitaxy:- Deposition of a single crystalline structure on a single crystalline substrate.

Homoepitaxy:- Deposition of single crystalline thin film on a single crystalline substrate of same material as of thin film.

Heteroepitaxy:- Two different materials for thin film and substrate.

**Nucleation**

Film growth starts with nucleation of different particles on the top of a substrate. Elementary processes can be well understood and visualized by considering atoms, ions or molecules in a plasma phase hitting top surface of the substrate.

The initial force behind nucleation is adsorption. Particles are bound by van-der-walls forces during adsorption. When particles forms chemical bond with substrates, the process is called as chemisorption. Desorption of particles is also possible during the process. Nucleation takes place agglomerated on surface of a substrate and forming chemical bonds with each other on substrate, as shown in figure 1.4.1.

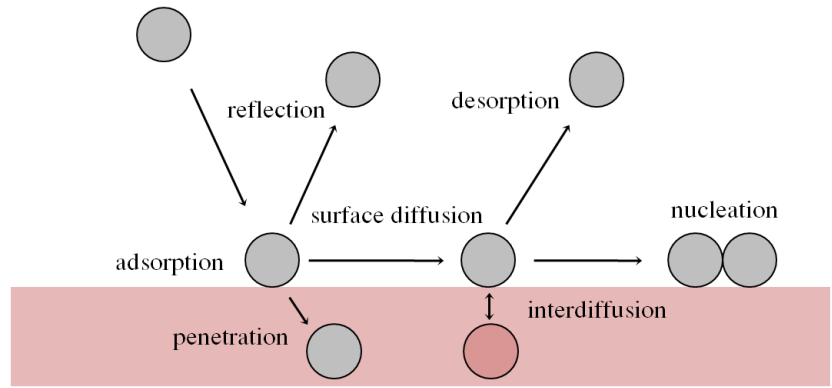

Figure 1.3.1. Schematic of initial process during thin film growth.

**Growth Modes**

Growth of nuclei is classically described by the interplay of surface energies of the film $\gamma_F$ and substrate $\gamma_S$ as well as interface energy between film and substrate $\gamma_I$ and the misfit energy $E_{mis}$. Growth modes for a thin film are of three types. The three modes are schematically depicted in figure 1.4.2.

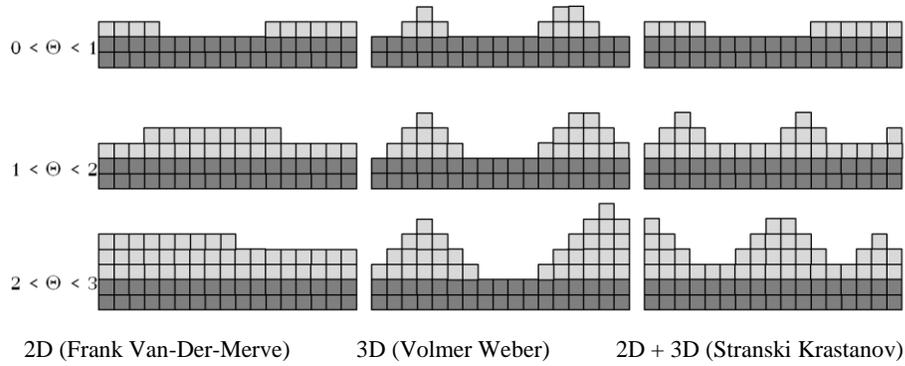

Figure 1.3.2. Schematic representation of different growth modes. Θ is the coverage of film material in units of monolayers

The three different possible cases are:

1. $\gamma_S > \gamma_F + \gamma_I$ ........................................ (1)

In this case film grows smoothly on the substrate and gets fully wetted by the film material. In this growth mode consecutive layers grows one by one. This is the ideal case for thin film growth. Considered as 2D growth.

2. $\gamma_S < \gamma_F + \gamma_I + E_{mis}$ .......................................... (2)

In this case formation of clusters occurs on top of the surface and considered as 3D growth.

3. $\gamma_S > \gamma_F + \gamma_I + E_{mis}$ ............................................ (3)

In some cases initially growth is of layer by layer i.e, 2D growth and then converts to 3D growth following the relation $\gamma_S < \gamma_F + \gamma_I + E_{mis}$.

# CHAPTER 2
# EXPERIMENTAL TECHNIQUES

## 2.1. Pulsed Laser Deposition

### 2.1.1 Basic principles of pulsed laser deposition

Pulsed laser deposition (PLD) is a universal method for fabrication of high quality and epitaxial films of metal, Semiconductors, insulators and organic materials. It utilizes the idea of using laser as a source for material ablation and was at first, tested in 1960's upon the creation of lasers [17]. The discovery of high-temperature superconductors in the late 1980's highlighted a special requirement of the PLD for thin film growth of different and complex materials, specially, multi-component oxides [18]. The PLD setup allows stoichiometric transfer of the target material towards the substrate and oxidation of thin film can be controlled by adjusting the background gas and its pressure. Recent development of Reflective High Energy Electron Diffraction (RHEED) technique allows the in-situ control on deposition of thin film. PLD technique is useful for atomic engineering of novel materials for advanced electronic applications and research [19].

In the PLD process, the material flux from the target is generated as the laser beam hits it, as shown in figure 2.1.1.1. Laser energy plays an important role during this process. If laser energy is low then the chemical composition of the material flux is dependent on the vapor pressure of target components resulting in thermal evaporation of the target. If the laser pulse fluence is higher than simple evaporation then material starts ablating. In ablation process, the energy absorbed by small volume is enough to ablate the entire volume and thus maintains the stoichiometry in the material flux or plume.

PLD setup which I have used during the first phase of my project work is as shown in figure 2.1.1.2. This PLD setup is having Nd:YAG laser to ablate target material.

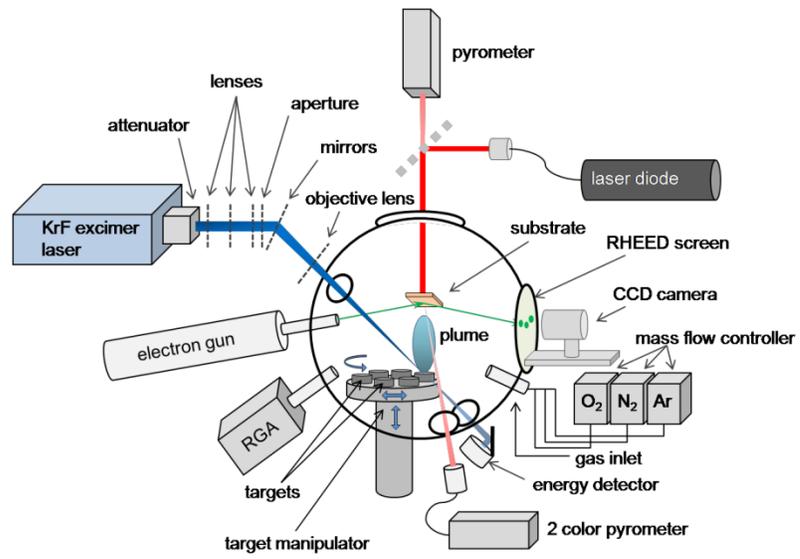

Figure 2.1.1.1. Schematic Representation of Pulsed Laser Deposition setup [Ref. 20]

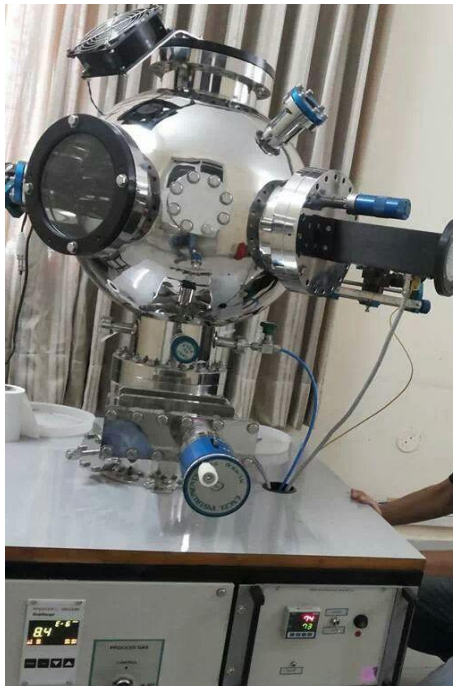

Fig. 2.1.1.2. PLD setup at IIT Patna

### 2.1.2. Excimer Lasers and Optics

According to Beer's law, the optical penetration depth $h$ of the laser beam into the target material depends on the ablation threshold $\varepsilon_o$ and laser fluence $\varepsilon$ in addition to the absorption coefficient of the material at a particular wavelength $a$ as [21].

$$h = a^{-1}(\varepsilon/\varepsilon_o) \quad \dots\dots\dots\dots\dots\dots\dots\dots\dots\dots\dots\dots\dots (4)$$

Ideally, all laser energy should be consumed by the target to ablate the material, but practically some part of energy is distributed in the target material through heat diffusion with a characterstic length of $l_{eff}$. This phenomenon cause heating of target material and may cause parallel evaporation of target material in addition to the ablation process. For an effective ablation process the condition $h < l_{eff}$ should be achieved.

This much amount of energy (~ 3J) is achieved by using excimer lasers operating at 193 nm (ArF), 248 nm (KrF), and 308 nm (XeCl) with a pulse duration of less than 30ns.

Optics includes a focusing convex lens, an aperture and an entrance window. The role of aperture is to define the shape of laser pulse and to ensure homogenous energy distribution of the laser. In general the efficiency of a optical system is around 30%.

### 2.1.3. PLD Targets

Oxide materials are pressed and sintered to make pellets which can be then used as a source to deposit a thin film by absorbing energy from the laser. A good pellet should have a density above 80% of the single crystal value to avoid microscopic particle ejection from the target on application of high energy lasr beam. Normally the stoichiometry of the target should be same as that of the film to be deposited but it is not a must requirement to achieve a desired phase in to be deposited thin film. Single crystals can also be used as targets but they should absorb laser energy sufficiently to make ablation possible. This technique also works with metal targets but needs some background gas such as oxygen. For oxide materials having low absorption coefficient such as MgO, this technique is worth to use.

Every time laser pulse hits the target material it ablate a small fraction of volume. The amount of ablated material is proportional to the laser fluence, density of the pellet and the beam area on the target surface. If the laser energy is inhomogenous it may cause ablation and evaporation simultaneously which can cause off stoichiometry in the thin film. Therefore, homogenous distribution of energy throughout the laser pulse is a must. The plume generated due to ablation of material is symmetric to the surface of pellet and in the direction of normal to the sample surface. The amount of the ablated material is a $\cos(\Phi^n)$ of the function of angle $\Phi$ from the normal to the target surface [22]. An asymmetric plume may generate if the laser hits the same area simultaneously. To avoid such condition one needs to move the spot along the substrate area and that can be done by rotating and sweeping the substrate simultaneously. Ideally target surface should be polish after every deposition but this is nor experimentally quite efficient but still after few deposition target should be polish to expose stoichiometric material from the pellet. Continuous hitting of laser may cause non stoichiometry of the top surface of the target material.

### 2.1.4. Vacuum and Gas Supply Unit

Deposition of a thin film is possible in vacuum as well as some gas environment such as nitrogen, oxygen, argon, helium, forming gas etc. oxygen generally introduced during the growth of oxide thin films and parameters like SCCM and pressure are the most important parameters in such case. This controls the oxidation of atomic species in the plume. Generally, molecular oxygen is used however atomic as well as ozone can also be used for such depositions. Another important task of background gas is to control the kinetic energy of the atomic particles in the plume which if not controlled can cause defects in the film. This can reduce the energy of particles less than 1 eV . For example the mixture of argon and oxygen with a pressure upto 1 mBar can be used to grow films of very good structural quality.

### 2.1.5. Substrate Heating System

After the material gets ablated from the target in the form of plume and has been delivered to the substrate and then nucleation and migration of species on the substrate occurs to start film growth. Structural quality of the deposited film, nucleation, cluster size, presence of defects, growth mechanism are the most important parameters characterizing atomic species migration on the film surface during the nucleation process, which determined by target to substrate distance, excimer laser fluence and thermodynamic conditions like gas pressure, atomic gas, and most importantly temperature of substrate.

Substrate temperature is one of the most important parameter to set during a PLD growth. Low temperature may result into a amorphous or polycrystalline thin films. In general for a high quality crystalline thin films, that should be greater than $500^o$ C so that the atomic species has sufficient mobility to nucleate and form a ordered structure. Typical temperature range for oxide materials during PLD growth is between $500^o$ C to $900^o$ C. Substrate heating can be done by gluing a substrate on a metal substrate holder and then heat it to attain the required temperature. Different methods are available for heating, like resistive heating, infrared radiation lamp and a modern technique like use of laser diode. The output of laser diode can coupled to PLD chamber with the help of UHV sealed optical window and IR beam is then focused on the back of the substrate i.e, metal substrate holder. This allows a very local heating keeping the environment at room temperature. local temperature up to $1100^o$ C is easily achievable. The temperature is controlled by thermocouple and/or infrared pyrometer which works on feedback loop systems. The advantage of using this method of is heating is, overall compatibility with oxygen and other gas environment allowing a longer life. however, other methods like resistive heating suffers due to oxidation of heating element with time. The only disadvantage of using this laser is the limited area which is due to the limited spot size of the laser beam. It is generally difficult to attain constant temperatures over large surfaces.

## 2.2. Reflection High Energy Electron Diffraction (RHEED)

### 2.2.1. Introduction to RHEED

Deposition rates up to $10^5$ nm/s are achievable with PLD. This rate is several order higher than the rate which can be achieved by other physical vapor deposition techniques. Generally, pulse counting and time duration can be used to control the deposition rate and film thickness in PLD process [23]. Recently, reflection high energy electron diffraction (RHEED) adopted for monitoring the in-situ growth of thin film in high background pressure of oxygen. It is an important tool for in-situ monitoring of surface atomic structure of the film during the growth. In RHEED, high energy electron beam (10-50 keV) diffract from the the film surface at grazing angle below $5^o$ due to interaction with few mono layers of the film. The diffraction pattern is then displayed on phosphor screen which can be used to analyze the structure and orientation of substrate as well as film. A vacuum better than $10^{-6}$ mbar is required for effective generation of electron beam. This requires vacuum pumps to achieve such low pressures. The pattern on a phosphor screen is captured by fast cam coder and then and transfer to a computer display.

### 2.2.2. In-situ Growth Monitoring by RHEED

RHEED can be used to monitor intensity variation in-situ and to monitor diffraction pattern on the screen. Bragg's equation explains the diffraction of waves through periodic lattice , in reciprocal space. $k_i$ is the incident wave vector and $k_{(h,k)}$ is the diffracted wave vector by the reciprocal lattice vector $G_{(h,k)}$.

$$k_{(h,k)} - k_i = G_{(h,k)} \quad \text{............................... (5)}$$

The magnitude for wave vector of high energy electron is given by

$$k_i = \frac{1}{\hbar}\sqrt{2m_0 E_{kin} + \frac{E_{kin}^2}{c^2}} \quad \ldots\ldots\ldots\ldots\ldots\ldots (6)$$

Where, $E_{kin}$ is kinetic energy of electron, $m_o$ is the rest mass of the electron, $c$ is speed of light. The wave vector of particles (electrons) accelerated by 10- 50 KV is very high, so that many diffraction conditions may be fulfilled at the same time. Since only few top most layers contribute to the diffraction pattern , the resulting reciprocal space is to be built by so called truncation rods or reciprocal lattice rods of the surface structure instead of reciprocal-lattice points. this leads to diffraction in beam direction and perpendicular to the beam direction. The basic principle of electron diffraction is as shown in Figure 2.2.1.1.

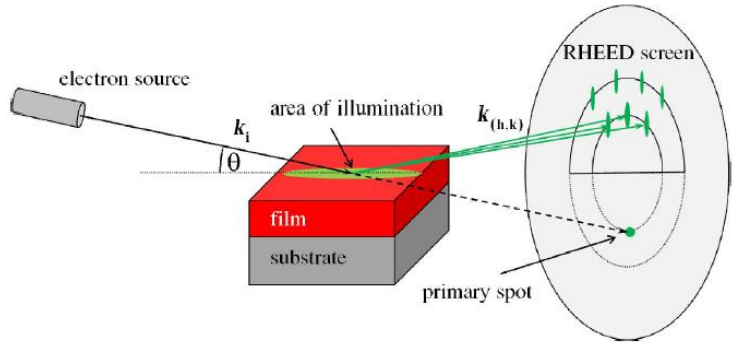

Figure 2.2.1.1. Principle of electron diffraction at crystal surface [ Ref. 20].

The RHEED image can also give information about morphology and surface structure. One can easily differentiate between different growth modes by just observing the RHEED pattern. Schematic of which is shown in Figure 2.2.1.2.

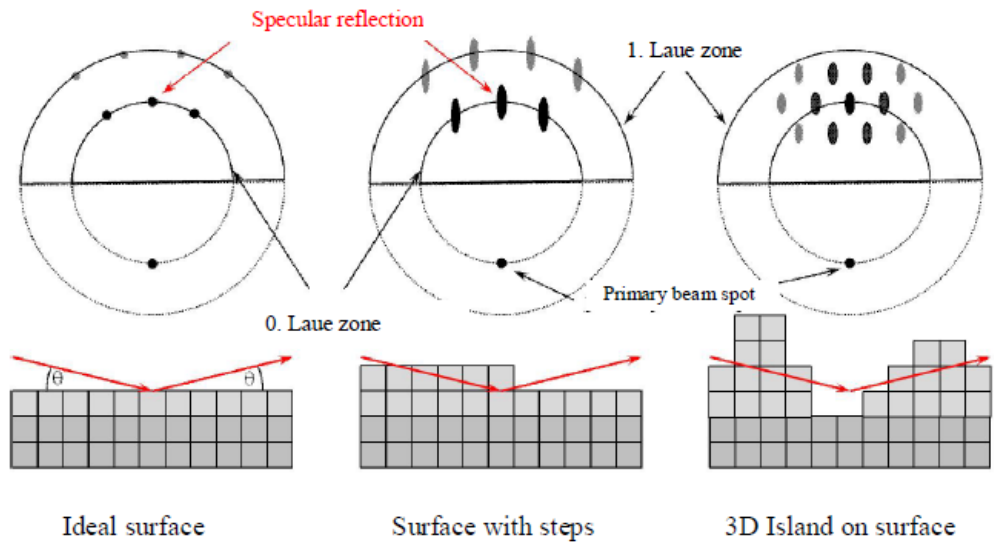

Figure 2.2.1.2. Schematic representation of surface morphologies and the resulting RHEED pattern. [Ref. 20]

## 2.3. Crystal Structure Analysis

### 2.3.1. X-Ray Diffraction

X-Ray Diffraction (XRD) is an analytical technique used for phase identification of crystalline materials and capable of giving information upto unit cell dimensions. Finely grinded powder, Bulk or even thin films can be studied [24-26].

This technique is based on constructive interference of monochromatic X-Rays when diffracted from a crystalline sample. These X-rays are generated by a cathode ray tube, then filtering is dine to produce monochromatic X-ray radiation, then collimated to concentrate, and then directed towards the sample. Constructive interference occurs when conditions satisfy Bragg's Law.

$$n\lambda = 2d \sin\theta \quad \text{.................. (7)}$$

Now, These diffracted X-ray beam first detected, then processed and counted. By scanning the available sample through the whole range of $2\theta$ angles, all diffraction peaks can be found out in case of powder sample due to availability of all random directions in powder sample. In case of thin film XRD this can be done by specifying the diffraction angle according to the substrate.

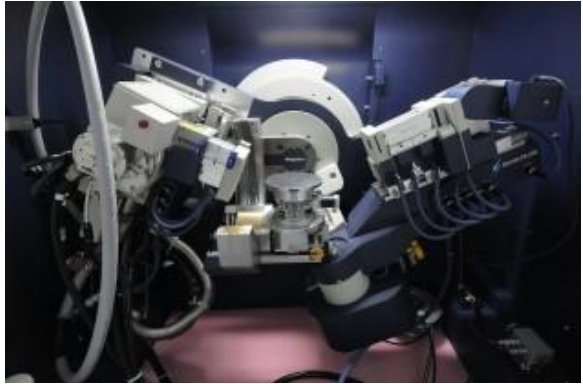

Figure 2.3.1.1. XRD at ATFT group, TU Darmstadt

**2.3.2. X-ray Reflectivity (XRR)**

X-ray reflectivity also known as X-ray reflectometry (XRR), is a surface sensitive technique used in material science, Physica, chemistry and fabrication technology to characterize thin films and heterostructures.

In this technique, X-ray beams reflected from a flat or smooth surface and then one measure the intensity of reflected beam in the specular direction. Deviation from the predicted value occur and can be used to analyze density profile, thickness [27] and roughness of the sample.

$$R(Q)/R_F(Q) = \left| \frac{1}{\rho_\infty} \int_{-\infty}^{\infty} e^{iQz} \left( \frac{d\rho_e}{dz} \right) dz \right|^2 \quad\ldots\ldots\ldots\ldots (8)$$

Equation (8) explains the basic mathematics behind the XRR technique. Where, $P_e(z)$ is the average electron density, $R(Q)$ is the reflectivity, $Q = 4\pi \sin(\theta)/\lambda$, $\lambda$ is the wavelength, $\rho$ is the density deep within the material.

## 2.4. Electron Microscopy

### 2.4.1. Field Emission Scanning Electron Microscopy (FE-SEM)

Field emission microscopy is an analytical technique used in physics, Chemistry and material Science to investigate surface structures and their respective properties. It was invented by Erwin Wilhelm Muller in 1936. It was first of the technique to reach atomic resolution.

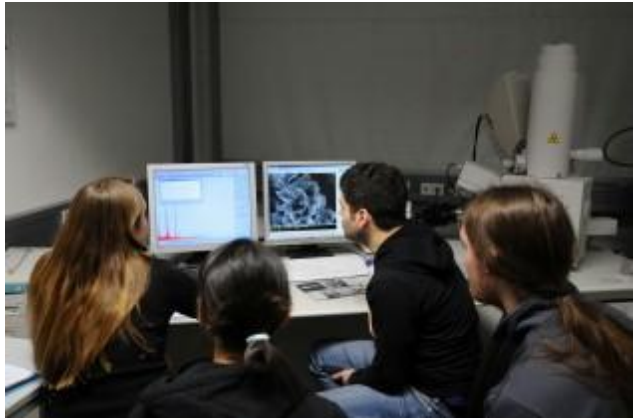

Figure 2.4.1.1. High Resolution Scanning Electron Microscopy (HR-SEM) at ATFT Group, TU Darmstadt

In this technique, electron beam generated by electron gun scans the sample surface and interacts with the atom at the specimen surface then genrate signals. These signals information about the specimen surface topography and morphology. Backscattered electrons gives information about the atomic number of species in the sample [28].

## 2.4.2. Energy Dispersive X-ray Analysis

Energy Dispersive X-ray Spectroscopy (EDS) or Energy Dispersive X-ray Analysis (EDXA), is an analytical technique used for the chemical characterization and elemental analysis of sample [29]. The characterization capabilities lies on the principle that every element has a unique atomic structure with unique X-ray emission spectra. To stimulate emission of X-rays from sample, a high energy beam of particle is bombarded on the sample, Such as electrons. The incident particle beam excite an electron from inner shell and creating a hole. Then, the electron from outer shells fills the hole and the difference energy is emitted in the form of X-rays. The number and energy of X-rays is then measured by an energy dispersive spectrometer.

## 2.5. Superconducting Quantum Interference Device (SQUID) Magnetometry

SQUID is a very sensitive magnetometer used to measure extremely small magnetic fields. It is having superconducting loop with josephson junction.

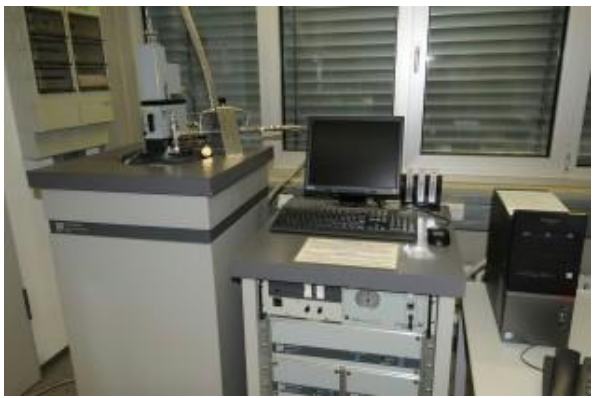

Figure 2.5.1. SQUID setup at ATFT group, TU Darmstadt

Magnetic field as low as 5 aT (5*10E-18 Tesla) can be measured using SQUID. SQUID magnetometer consists of magnet which produce massive local field, a super conducting detection coil which inductively linked to the sample and magnetic shield around the SQUID setup. Detection coil measures the changes in local field. SQUID gives Field cooled (FC) and Zero-Field Cooled (ZFC) plots of M-T measurement at temperature range from room temperature to 5K, and M-H plot at any temperature in the above specified range [30].

# CHAPTER 3
# RESULTS AND DISCUSSIONS

The target of LaMnO$_3$ (LMO) was prepared using solid state synthesis route. Silver glue was used to glue the target on target holder of the pulsed laser depostion (PLD). STO (001) substrates have been used to deposit LMO thin films using PLD. Parameters like film thickness, laser fluence, pulse frequency, background gas, substrate temperature have been varied to investigate the effect of these on magnetic properties of LMO thin films.

In the following section results of pellet preparation, X-ray diffraction (XRD), Energy dispersive spectroscopy(EDS), Scanning electron microscopy (SEM), Super conducting quantum interference device(SQUID) are shown and discussed.

## 3.1. Bulk LaMnO$_3$

Lanthanum Mangenese Oxide (LaMnO$_3$) was prepared by using conventional solid state synthesis route [31]. La$_2$O$_3$ and Mn$_2$O$_3$ were used as precursors. The balanced chemical reaction is given in equation (9) and prepared pellet as shown in figure 3.1.2.

$$0.5 \text{ La}_2\text{O}_3 + 0.5 \text{ Mn}_2\text{O}_3 \implies \text{LaMnO}_3 \quad \text{................(9)}$$

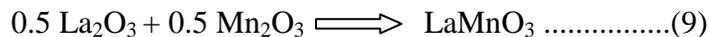

The precursors were mixed in stoichiometric amount and then finely grinded using mortar and pestle. Two step calcination process was followed to make a single phase LMO target. The XRD pattern of calcined powder confirms the LMO phase, as shown in figure 3.1.2. The results of each step of pellet preparation ia as shown in Table 3.1.1.

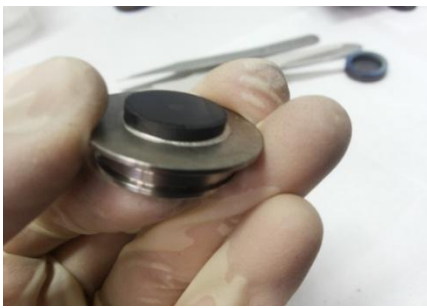

Figure 3.1.1. LaMnO$_3$ pellet mounted on target holder.

Table 3.1.1. Results of 'pellet preparation process'

| | |
|---|---|
| Total powder weight | 5 gm's |
| La$_2$O$_3$ Weight | 3.368000 gm's |
| Mn$_2$O$_3$ Weight | 1.632000 gm's |
| La$_2$O$_3$ Pre Calcination | 300$^o$ C for 12 Hours |
| Grinding time for precursors | 1 Hour 15 Minutes |
| Amount recovered after grinding | 4.9862 gm's |
| First calcination | 900$^o$ C (5$^o$ C/min) for 12 Hours |
| Grinding (Second time) | 01 Hour |
| Second calcination | 1300$^o$ C (5$^o$ C/min) for 12 Hours |
| Pellet preparation(using hydraullic press) | 4 Tonns for 30 Seconds |
| Pellet sintering | 1300$^o$ C for 12 Hours |
| Sintered pellet weight | 2.8974 gm's |
| Pellet thickness | ~ 2.7 mm |
| Pellet diameter | 15 mm |
| Pellet Density | ~ 6 mg/(mm)$^3$ |

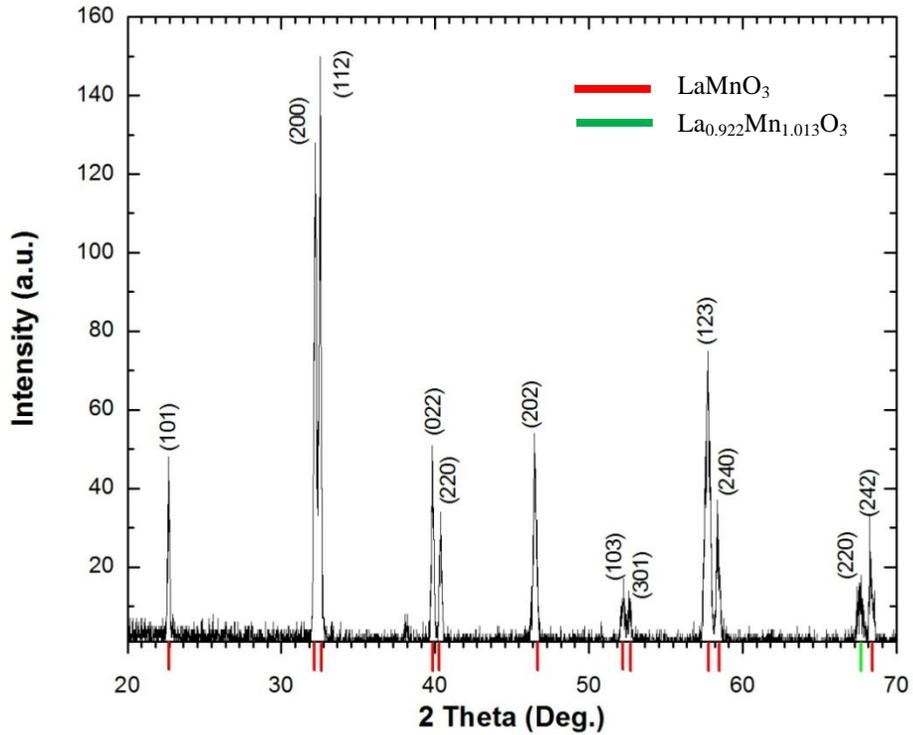

Figure 3.1.2. XRD pattern for LMO pellet

## 3.2. Scanning Eletron Microscopy Results

The SEM images of target surface are as shown in figure 3.2.1. Before analyzing pellet using SEM, the surface have been polished using a sand paper. There were few small cracks on the surface of sintered pellet, however, after polishing the surface becomes smooth. The target is highly dense and can be used as target for PLD system.

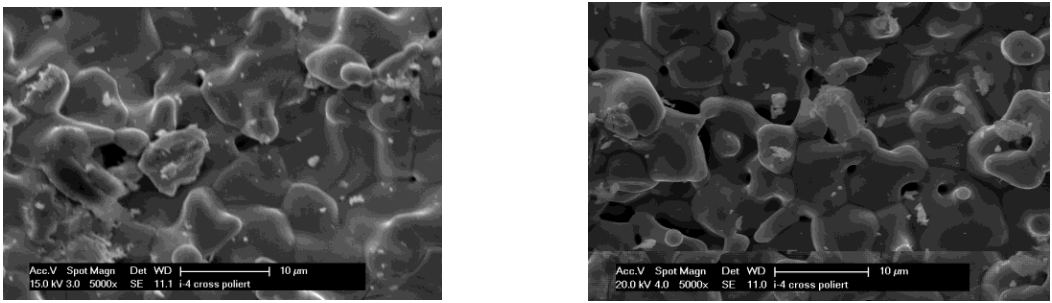

Figure 3.2.1. SEM images of LMO pellet

## 3.3. Energy Dispersive Spectroscopy (EDS) Results

Energy dispersive spectroscopy results are as shown in figure 3.3.1. Elemental composition of LMO pellet is given in table 3.3.1. Ideally, La and Mn composition should be 50% - 50%. In our case composition is very much close to the chemical formula i.e, La::Mn is 49.54::50.46

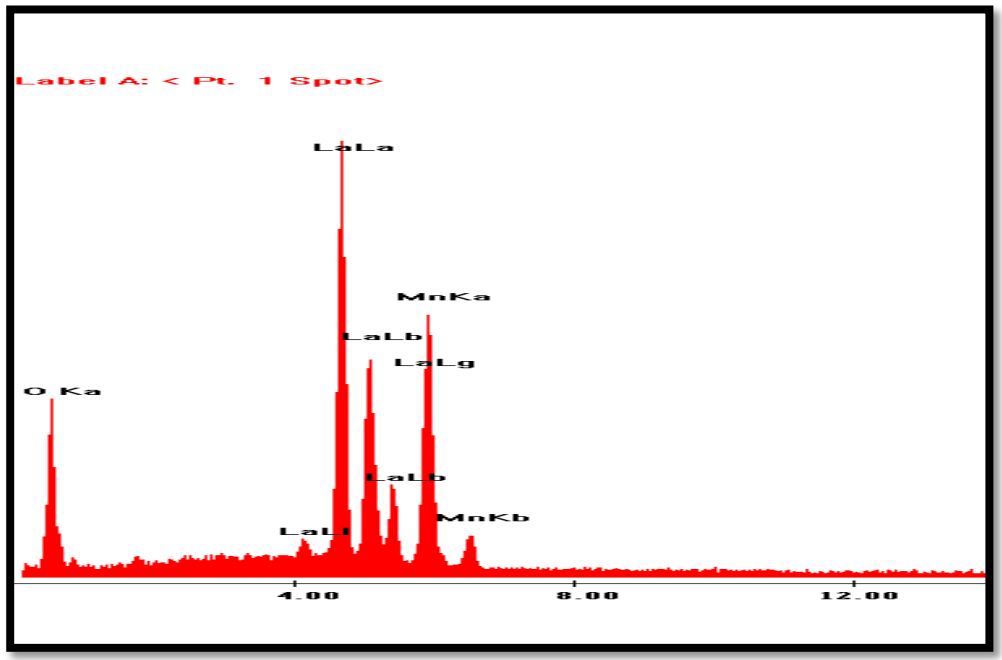

Figure 3.3.1. EDS of LMO pellet

Table 3.3.1. Elemental percentage of Lanthanum and Manganese in prepared target.

| Element | Wt% | At% | K-Ratio | Z | A | F |
|---|---|---|---|---|---|---|
| LaL | 71.28 | 49.54 | 0.7256 | 0.9584 | 1.0388 | 1.0225 |
| MnK | 28.72 | 50.46 | 0.2180 | 1.0853 | 0.6995 | 1.0000 |
| Total | 100.00 | 100.00 | | | | |

### 3.4. Fabrication Parameters

Fabrication parameters for all deposited LMO thin films are as shown in Table 3.4.1. and characterization and results are discussed in the following sections.

Table 3.4.1. Fabrication parameters for LMO thin films

| Film No. | Pressure (m Torr) | Gas | Fluence (J/cm$^2$) | Frequency (Hz) | Pulse Shots | Deposition Temperature ($^o$C) | Remarks |
|---|---|---|---|---|---|---|---|
| 01 | SUBSTRATE DISCARDED | | | | | | |
| 02 | 10 | O$_2$ | 1 | 4 | 3000 | 650 | |
| 03 | 0.1 | O$_2$ | 1 | 4 | 3000 | 650 | |
| 04 | 0.01 | O$_2$ | 1 | 4 | 3000 | 650 | |
| 05 | 0.01 | O$_2$ | 1 | 4 | 1000 | 650 | |
| 06 | 1 | Ar | 1 | 2 | 1000 | 650 | |
| 07 | 1 | Ar | 1 | 2 | 1000 | 600 | |
| 08 | 1 | Ar | 1 | 2 | 1000 | 650 | |
| 09 | 1 | Ar | 1 | 4 | 1000 | 650 | |
| 10 | 1 | Ar | 1 | 2 | 1000 | 650 | Vac Cooled |
| 11 | 1 | Ar | 1 | 4 | 1000 | 650 | Vac Cooled |
| 12 | 6.7E-5 | Vac | 0.75 | 2 | 1000 | 650 | |
| 13 | 6E-5 | Vac | 1 | 2 | 1000 | 650 | |
| 14 | 2.3E-5 | Vac | 0.5 | 2 | 1000 | 650 | |
| 15 | 4.4E-5 | Vac | 0.5 | 2 | 750 | 650 | |
| 16 | 0.1 | Ar | 0.75 | 2 | 1000 | 650 | |
| 17 | 6.4E-5 | Vac | 0.7 | 2 | 1000 | 650 | |
| 18 | 5.2E-5 | Vac | 0.6 | 2 | 1000 | 650 | |
| 19 | 4E-5 | Vac | 0.75 | 2 | 1000 | 650 | |
| 20 | 8E-5 | Vac | 0.75 | 2 | 3000 | 650 | |
| 21 | 6.5E-5 | Vac | 0.5 | 2 | 4000 | 650 | |
| 22 | 0.1 | Ar | 0.75 | 2 | 4000 | 650 | |
| 23 | 0.1 | Ar | 1 | 2 | 4000 | 650 | |
| 24 | 0.1 | Ar | 0.5 | 2 | 4000 | 650 | |
| 25 | 0.1 | Ar | 0.6 | 2 | 4000 | 650 | |
| 26 | 0.1 | Ar | 0.7 | 2 | 4000 | 650 | |
| 27 | 0.1 | Ar | 0.7 | 2 | 4000 | 650 | |
| 28 | 1 | Ar | 0.5 | 2 | 3000 | 650 | Bad Quality |
| 29 | SUBSTRATE DISCARDED | | | | | | |
| 30 | 3.1E-5 | Vac | 0.4 | 2 | 4000 | 650 | |

LMO thin films were deposited in oxygen, Argon and vacuum background on STO(001) substrates with TiO$_2$ layer termination [32].. All the important parameters of a PLD deposition process and possible LMO thin film growth conditions are as shown in figure 3.4.1. All films were deposited using same DCA-PLD setup.

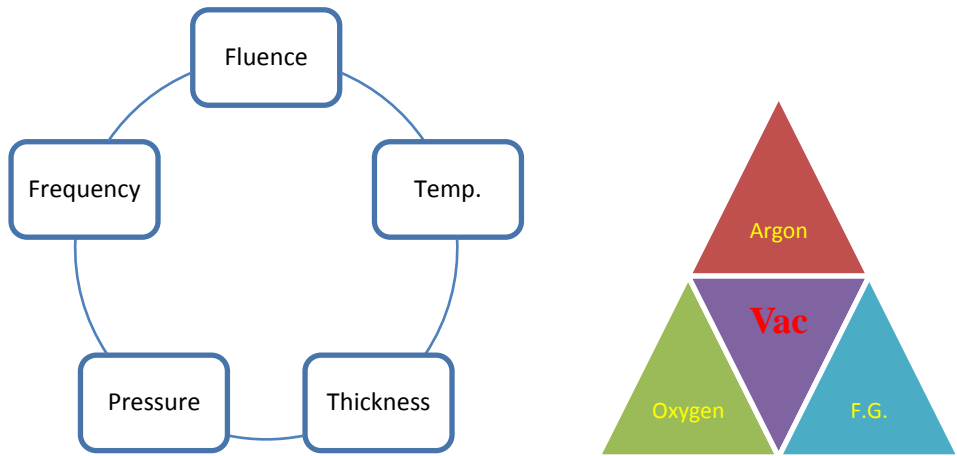

Figure 3.4.1. PLD parameters and background gas.

### 3.4.1. Sample Preparation

For each deposition, sample preparation was carried out in the similar manner. First a $TiO_2$ terminated STO (001) substrate is cleaned by using acetone and ethanol to remove dirt particles, if any. Silver paste is used to glue the substrate to the sample holder. Substrate holder then be heated for 10 minutes at a temperature of $80^oC$. After this sample holder is installed inside the load lock and afterwards transferred to the main chamber. All the steps are as shown in figure 3.4.1.1.

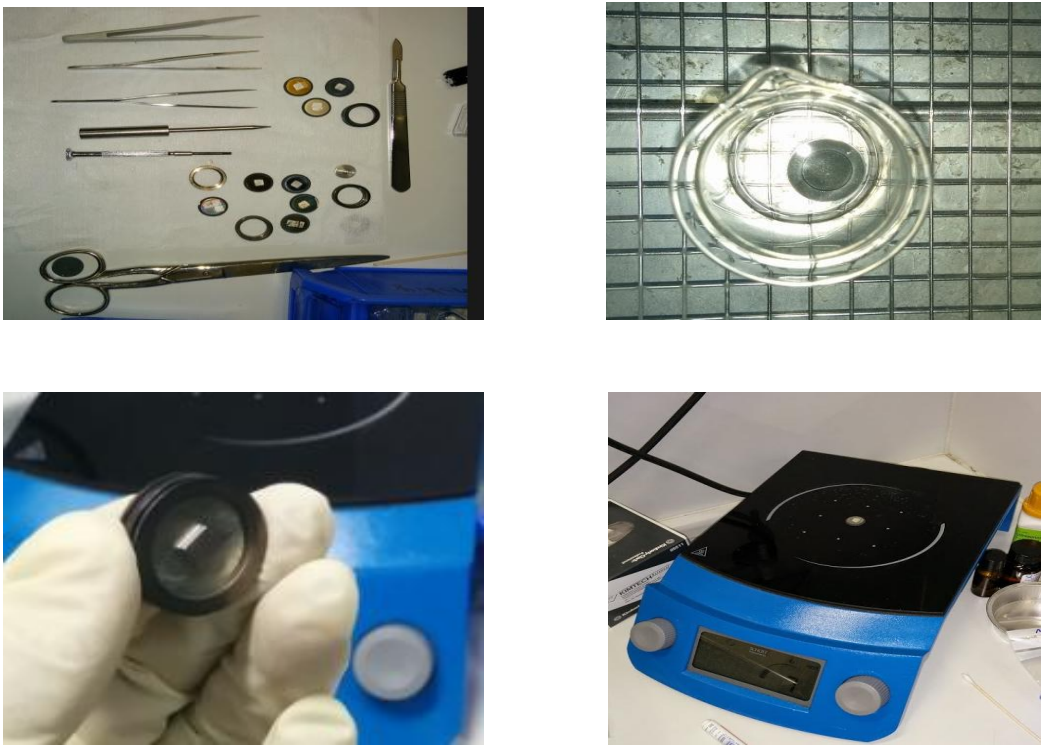

Figure 3.4.1.1. Steps for sample preparation

## 3.4.2. Parameters to be Analyzed for Every Deposition

There are many parameters and conditions to be take care of. A list of such parameters and conditions which are to be set and analyzed during the deposition are as shown in Table 3.4.2.1.

Target to substrate distance was fixed for all films and approximately equals to 40mm. Number of pulses for various films varied from 750 to 4000. Spot size of the laser beam was 3 mm$^2$. Before each deposition, energy check has been done. Target rotation was kept at 2 while target sweep at 0.27 for all films. Pressure inside the chamber was controlled by local/auto mode. Emissivity of the substrate holder was set as 0.6. RHEED voltage was fixed at 50KV and current was varied according to the real time intensity of RHEED pattern. Substrate heating was done using IR laser and KrF excimer was used to ablate target material with wavelength of 248nm.

Table 3.4.2.1. Parameters to be analyzed for every deposition

| Target-Substrate Distance | Pulse Frequency | Number of Pulses | Energy Detection | Laser Input Voltage | Target Rotation |
|---|---|---|---|---|---|
| Coolant Temperature | RHEED Power | Emissivity | Pressure Control Mode | Gas Flow Valves | Target Sweep |

## 3.5. Reflection High Energy Electron Diffraction(RHEED) Analysis

RHEED is a technique used to characterize surface of crystalline material. Figure 3.5.1. shows the intensity pattern of diffraction spots versus time of LMO thin film deposited in oxygen background. These oscillations are very much on the same level of intensity ,which confirms good crystalline quality of the deposited film i.e, layer by layer growth. At the end, due to thickness effect, RHEED oscillations vanishes. One can calculate the film thickness using this oscillations. Each oscillation corresponds to a unit cell growth. So, number of oscillation times the c-lattice constant gives the approximate film thickness.

Thickness Calculation for LMO 05:

Number of Pulses (p): 1000

Pulse Frequency (f) : 4

Therefore, Time (T) = (p/f) = 250 Seconds

Approx c-lattice constant = 0.4 nm

One oscillation duration (t) = 6.67 Seconds

Therefore, number of unit cells deposited = (T/t) = 37.5

Approximate Thickness (z) = (37.5 * 0.4) ~ 15.01 nm

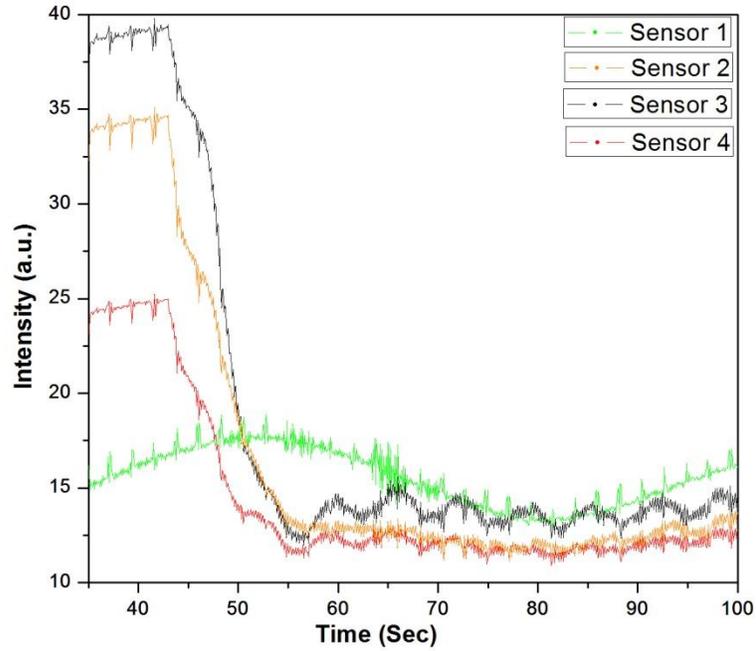

Figure 3.5.1. RHEED pattern for LMO 05

As shown in Figure 3.5.2, RHEED diffraction pattern from bare substrate and at the end of film growth is shown. Pattern confirms good surface quality of bare substrate, as shown in Figure 3.5.2 (a). As shown in RHEED oscillation that initially growth is layer by layer. However, at the end the growth mode changes and becomes layer plus island growth. Streaks confirms layer by layer plus island growth, as shown in Figure 3.5.2. (b).

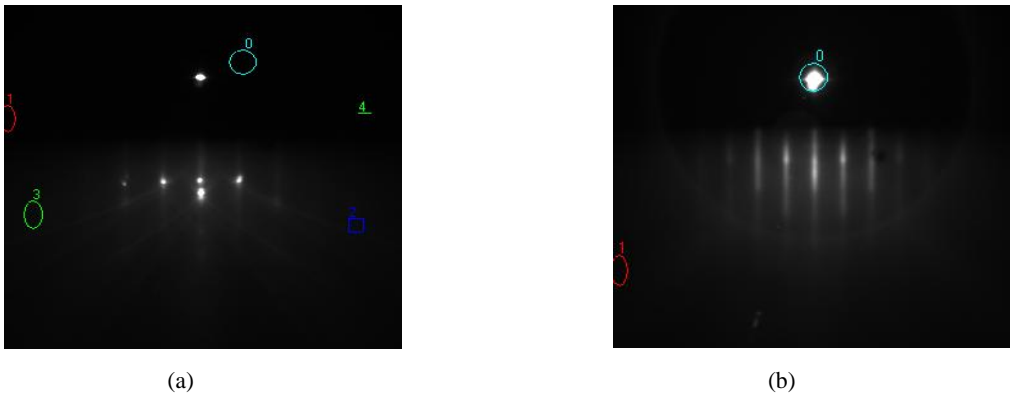

(a)            (b)

Figure 3.5.2. RHEED spots at different time (a) bare substrate (b) at the end of the deposition

## 3.6. X-ray Reflectivity (XRR) analysis

All the film thicknesses calculations were done using Rigaku's GIXRR fitting software. Film thickness, Film roughness and film density can be calculated using this software. Similar to other techniques this also has some error. Figure 3.6.1. shows XRR curve for sample LMO 05.

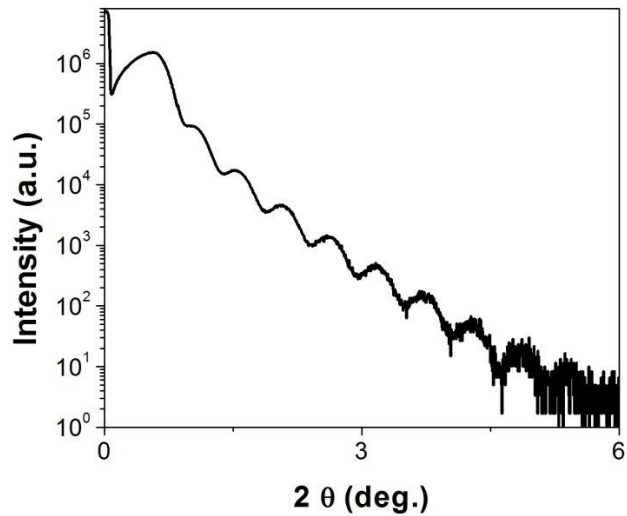

Figure 3.6.1. XRR curve for LMO 05

## 3.7. LMO Thin Films Grown in Oxygen

To get the desired LMO phase, few films were deposited in oxygen background. As per literature these were expected to be highly ferromagnetic [33-35]. To compare with the existing result, few LMO thin films were deposited at different oxygen pressure with all other parameters kept constant as per Table 3.7.1.1. Background pressure was varied from 10 mTorr to 0.01 mTorr.

### 3.7.1. Structural Properties of LMO Thin Films Grown in Oxygen Background

Structural properties of LMO thin films were studied using XRD θ-2θ scans and XRR fitting.

Table 3.7.1.1. Film deposition parameters for different backgroung gas pressure

| | |
|---|---|
| Fluence | 1 J/cm$^2$ |
| Frequency | 4 |
| Pulse shots | 3000 |
| Temperature | 650 $^o$C |
| Target to Substrate Distance | 40 mm |
| Background | O$_2$ |

After the deposition, XRD analysis was done for all the films. c-lattice constant and thickness were calculated using XRR. All these results are summarized in Table 3.7.1.2.

Table 3.7.1.2.. c-lattice parameter and film thickness of films grown in oxygen

| Film Number | Pressure (m Torr) | Thickness (nm) | c-lattice (A) |
|---|---|---|---|
| LMO 02 | 10 | 41.5 | 3.947 |
| LMO 03 | 1 | 47.8 | 3.980 |
| LMO 04 | 0.1 | 50.25 | 3.983 |
| LMO 05 | 0.1 | 15.69 | 3.978 |

XRD plots confirms the LMO phase in all the thin films, as shown in Figure 3.7.1.1. Temperature of 650$^o$C and fluence of 1J/cm$^2$ were good growth conditions to grow LMO.

All these films shows beautiful Laue oscillations. Laue oscillations for LMO 04 are as shown in Figure 3.7.1.2.

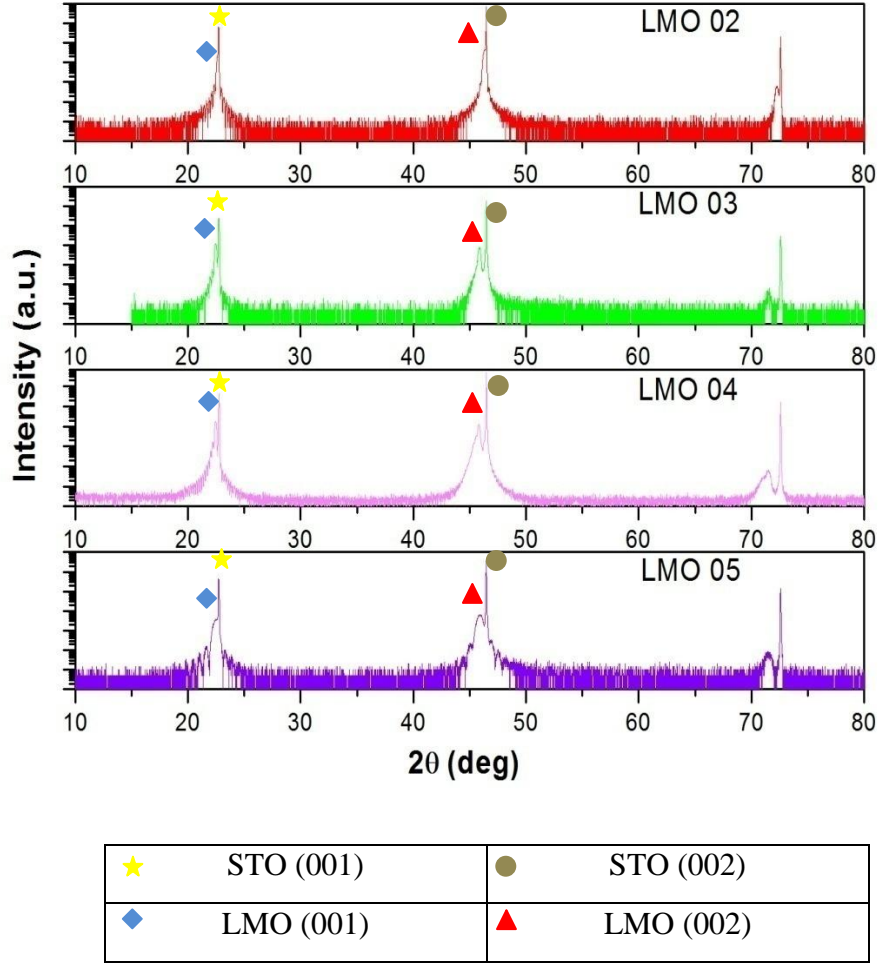

| ★ | STO (001) | ● | STO (002) |
| --- | --- | --- | --- |
| ◆ | LMO (001) | ▲ | LMO (002) |

Figure 3.7.1.1. XRD θ-2θ scan for LMO thin films grown in oxygen background

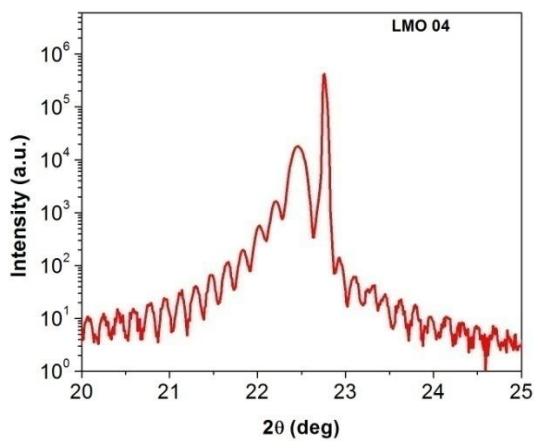

Figure 3.7.1.2. Laue oscillations in LMO 04

### 3.7.2. Magnetic Properties of LMO Thin Films Grown in Oxygen Background

Magnetic properties of LMO thin films grown in oxygen background were same as expected. These films were highly ferromagnetic with saturation moment($M_S$) grater than 2 $\mu_B$/f.u [33]. The curie temperature($T_c$) was in accordance with the reported values. The measured $T_c$ for oxygen grown films was around 140K [36]. Results are as shown in Table 3.7.2.1 and in Figure 3.7.2.1.

Table 3.7.2.1. Magnetic moment for LMO 03

| Film Number | Background Gas | Magnetic Moment |
|---|---|---|
| 03 | Oxygen | 2.2 $\mu_B$/f.u. |

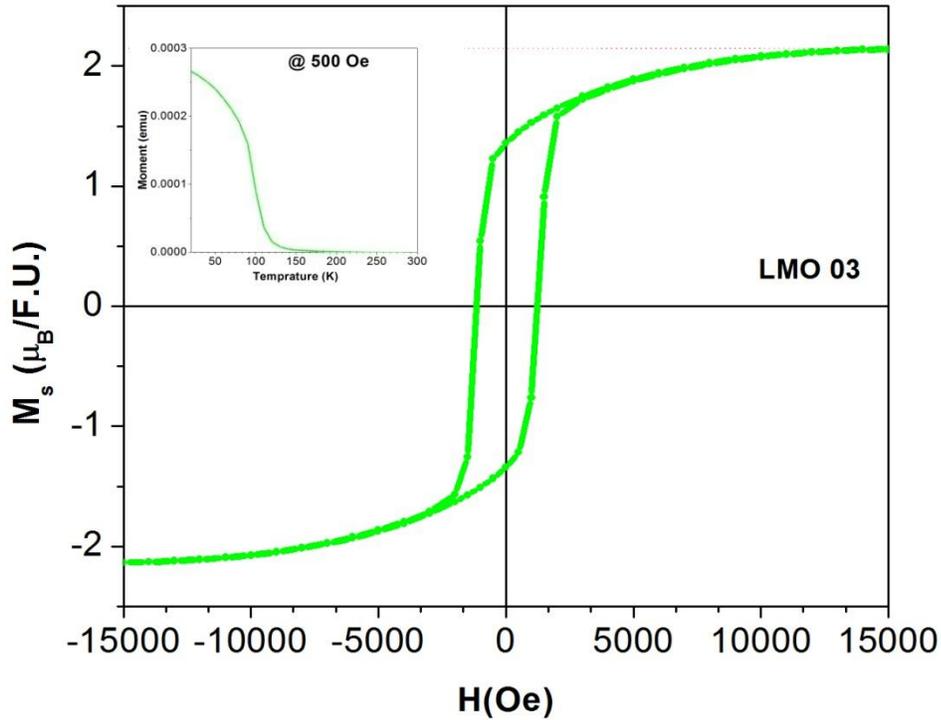

Figure 3.7.2.1. M-H Loop for thin film grown in oxygen background measured at 20K. The inset shows M-T curve measured at 500 Oe

The above results confirms the prescence of $Mn^{4+}$ ions in the film i.e, these films are oxygen rich. Double exchange between $Mn^{3+}$ and $Mn^{4+}$ leads to such high ferromagnetism[37-42].

## 3.8. LMO Thin Films Grown in Vacuum

LMO thin film growth is possible in vacuum. As there is no literature available on vacuum growth of LMO so, optimizing LMO growth in such conditions needs optimisation of each and every parameter. Most important parameters to be optimized are fluence, temperature, thickness.

### 3.8.1. Structural Properties of LMO Thin Films Grown in Vacuum

Most important thing for thin film deposition in vacuum is the 'flence threshold'. Higher fluence leads to too much reducing conditions. LMO thin films were deposited at different fluences with all other parameters kept constant as per Table 3.8.1.1. The results of XRD and XRR analysis are given in Table 3.8.1.2.

Table 3.8.1.1. deposition parameters for different fluences

| Pressure | Vacuum |
|---|---|
| Frequency | 2 |
| Pulse shots | 1000 |
| Temperature | 650 °C |
| Target to Substrate Distance | 40 mm |
| Background | Vacuum |

Table 3.8.1.2.. c-lattice parameter and film thickness of films grown in vacuum with different fluences

| Film Number | Fluence (J/cm$^2$) | Thickness (nm) | c-lattice (A) |
|---|---|---|---|
| LMO 13 | 1 | La$_2$MnO$_4$ | |
| LMO 12 | 0.75 | Mixed Phase | |
| LMO 17 | 0.7 | Mixed Phase | |
| LMO 14 | 0.5 | 9.84 | 4.026 |

The fluence scan confirms the fluence threshold for LMO growth to be 0.5 J/cm$^2$. Above this threshold fluence, some mixed phases appeared in the films and La$_2$MnO$_4$ at fluence 1 J/cm$^2$ or higher. XRD θ-2θ scans are as shown in Figure 3.7.1.1.

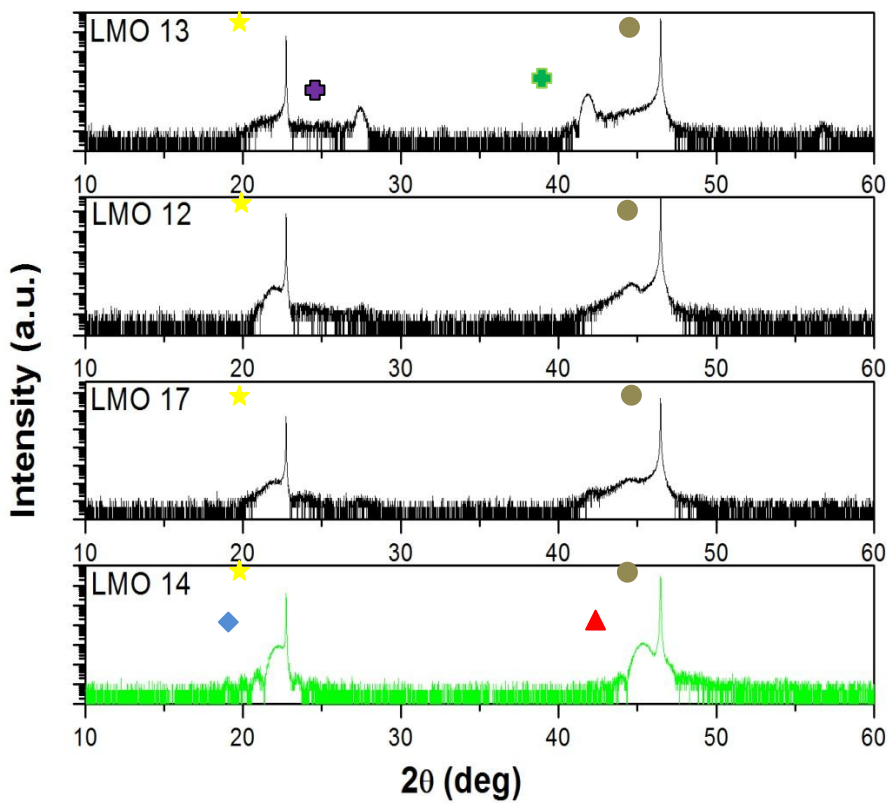

| ★ | STO (001) | ● | STO (002) |
| ♦ | LMO (001) | ▲ | LMO (002) |
| ✚ | La$_2$MnO$_4$ (001) | ✚ | La$_2$MnO$_4$ (002) |

Figure 3.8.1.1. XRD θ-2θ scan for LMO thin films grown in vacuum with different fluence.

After getting the fluence threshold, thickness scan was performed to find out the effect of thickness on magnetic properties of LMO thin films. Films were deposited with different pulses shots having other parameters kept constant as shown in Table 3.7.1.3. The results of XRD θ-2θ scan and XRR are as shown in Table 3.7.1.4 with XRD θ-2θ scan plots in Figure

Table 3.8.1.3. deposition parameters for different thicknesses

| Pressure | Vacuum |
|---|---|
| Frequency | 2 |
| Fluence | 0.5 J/cm$^2$ |
| Temperature | 650 °C |
| Target to Substrate Distance | 40 mm |
| Background | Vacuum |

Table 3.8.1.4. c-lattice parameter and film thickness of films grown in vacuum with different thicknesses

| Film Number | Pulses | Thickness (nm) | c-lattice (A) |
|---|---|---|---|
| LMO 21 | 4000 | 42.88 | 4.061 |
| LMO 14 | 1000 | 9.84 | 4.026 |
| LMO 15 | 750 | 7.68 | 3.977 |

Results clearly shows the LMO growth at 0.5 J/cm$^2$ fluence. The lattice constant increases with increase in the thickness and that concludes the presence of Mn$^{2+}$ ions and their increase in concentration as the thickness increase. However, the results are quite promising and ferromagnetic order reduced considerably. The results of magnetic measurements are discussed in the coming sections.

A lower fluence of 0.4 J/cm$^2$ (with all other deposition parameters similar Table 3.8.1.3) was also deposited and this leads to very good magnetic properties. XRD θ-2θ scan plot as shown in Figure 3.8.1.3.

Table 3.8.1.5. c-lattice parameter and film thickness of films grown in vacuum with fluence of 0.4 J/cm$^2$

| Film Number | Pulses | Thickness (nm) | c-lattice (A) |
|---|---|---|---|
| LMO 30 | 4000 | 33 | 4.016 |

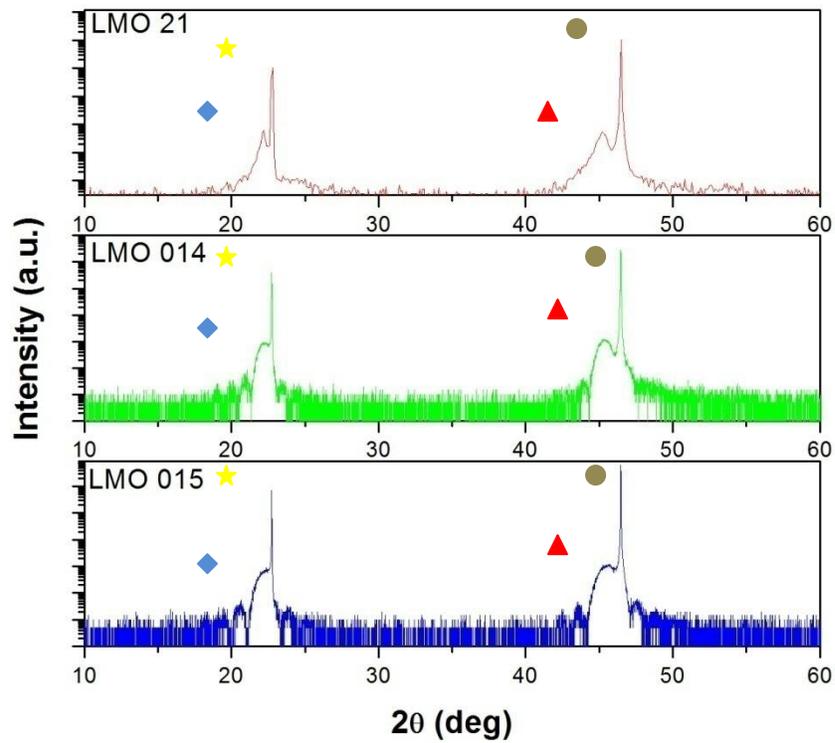

| ★ | STO (001) | ● | STO (002) |
| ◆ | LMO (001) | ▲ | LMO (002) |

Figure 3.8.1.2. XRD θ-2θ scan for LMO thin films grown in vacuum with different thickness.

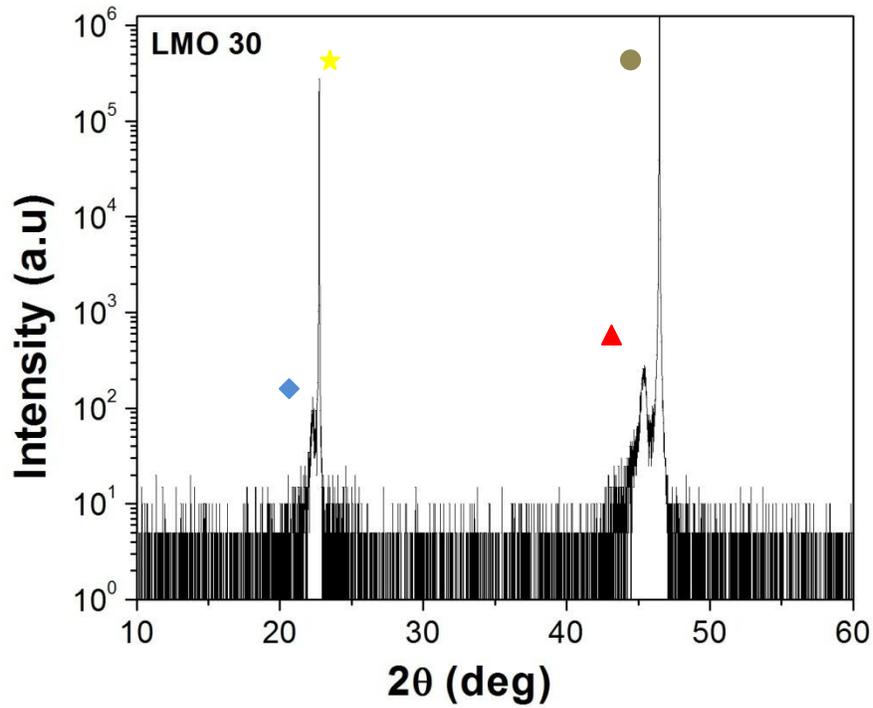

Figure 3.8.1.3. XRD θ-2θ scan for LMO thin films grown in vacuum with fluence 0.4 J/cm$^2$

The growth mode for LMO thin films on STO(001) under vacuum background, is cubic-on-cubic [43]. Phi scan was performed and result of which is shown in Figure 3.8.1.4.

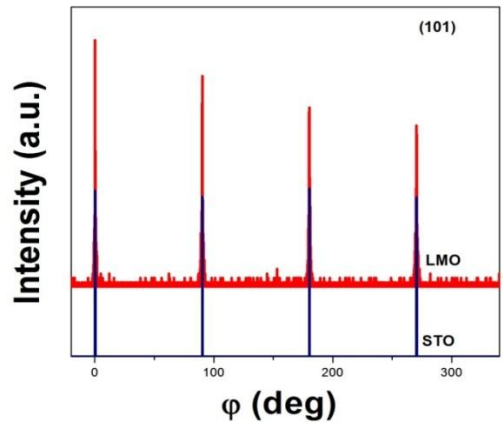

Figure 3.8.1.4. Phi-scan of vacuum grown LMO thin film in 101 plane

### 3.8.2. Magnetic Properties of LMO Thin Films Grown in Vacuum

Thin films grown in vacuum showed low magnetic saturation as compared to LMO thin films grown in oxygen background. Values of saturation moment are given in Table 3.8.2.1. The M-H loops for the LMO films with different thicknesses grown in vacuum are as shown in Figure 3.8.2.1. M-H loops experienced vertical hysterisis shift which is quite uncommon in single perovskite [44-49]. But as shown in figure 3.8.2.1. This shift increases with thickness. Most importantly, this is reproducible. However, it is not scientifically explainable why M-H loops experienced such huge shift. One possible reason could be the presence of pinned moments across the interface between film and substrate. M-T curves were taken at field of 500 Oe [50-52]. As shown in Figure 3.8.2.2, $T_c$ increases as thickness increases. and thicker film approaches the reported value of $T_c=140K$.

Table 3.8.2.1. Magnetic moment for LMO thin film with different thicknesses grown in vacuum

| Film Number | Pulses | Magnetic Moment |
|---|---|---|
| LMO 21 | 4000 | 0.35 $\mu_B$/f.u. |
| LMO 14 | 1000 | 0.28 $\mu_B$/f.u. |
| LMO 15 | 750 | 0.63 $\mu_B$/f.u. |

Further reduction of fluence leads to better magnetic properties. Figure 3.8.2.3 shows the M-H loop and M-T curve respectively. The deposition parameters for this LMO thin film are as given in Table 3.8.1.3 and 3.8.1.5. Magnetic saturation value is given in Table 3.8.2.2.

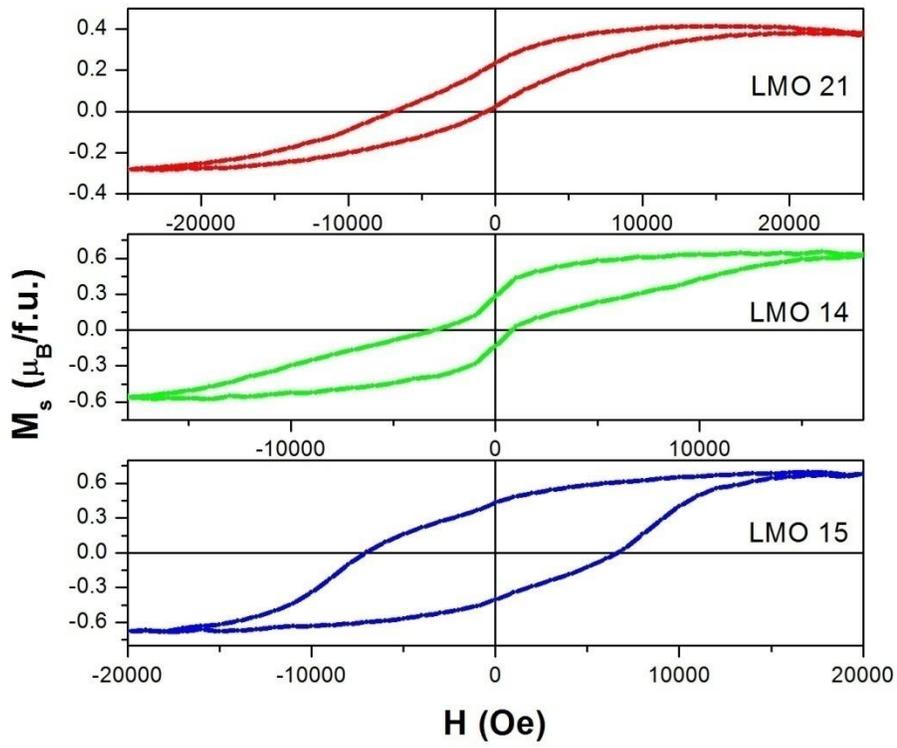

Figure 3.8.2.1. M-H loops for LMO thin film with different thicknesses grown in vacuum at 20K

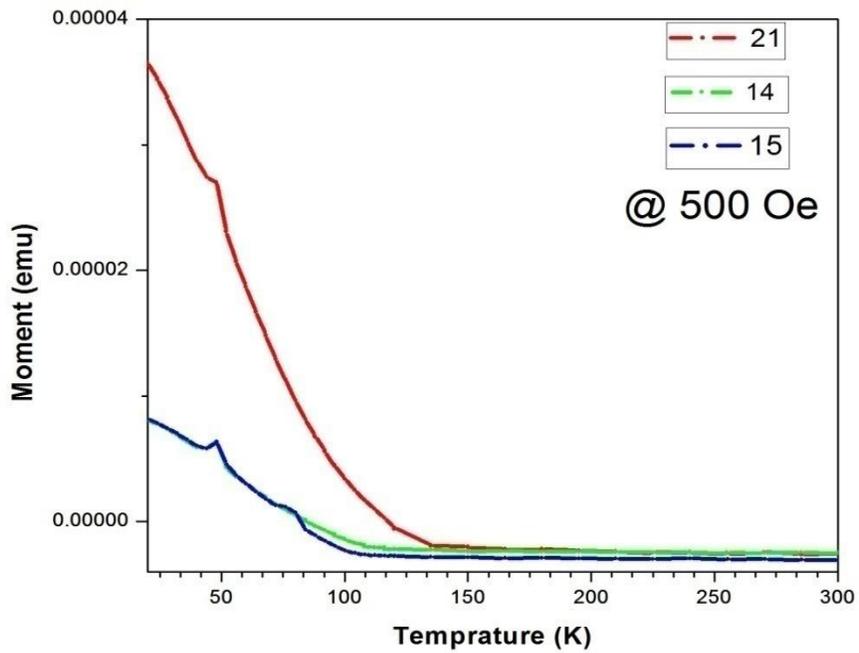

Figure 3.8.2.2. M-T curves for LMO thin film with different thicknesses grown in vacuum(measured at 500Oe

Table 3.8.2.2. Magnetic moment for LMO 30

| **Film Number** | **Pulses** | **Magnetic Moment** |
|---|---|---|
| LMO 30 | 4000 | 0.275 $\mu_B$/f.u. |

As discussed earlier, M-H loops experienced shifts in vertical axis. To investigate, whether this is an experimental error, machine error or it's came from film itself, three measurements were performed on the same sample with Field Cooled (FC) with +1 Tesla and -1 Tesla field with third one a Zero Field Cooled(ZFC) measurement. Results are as shown in Figure 3.8.2.4.

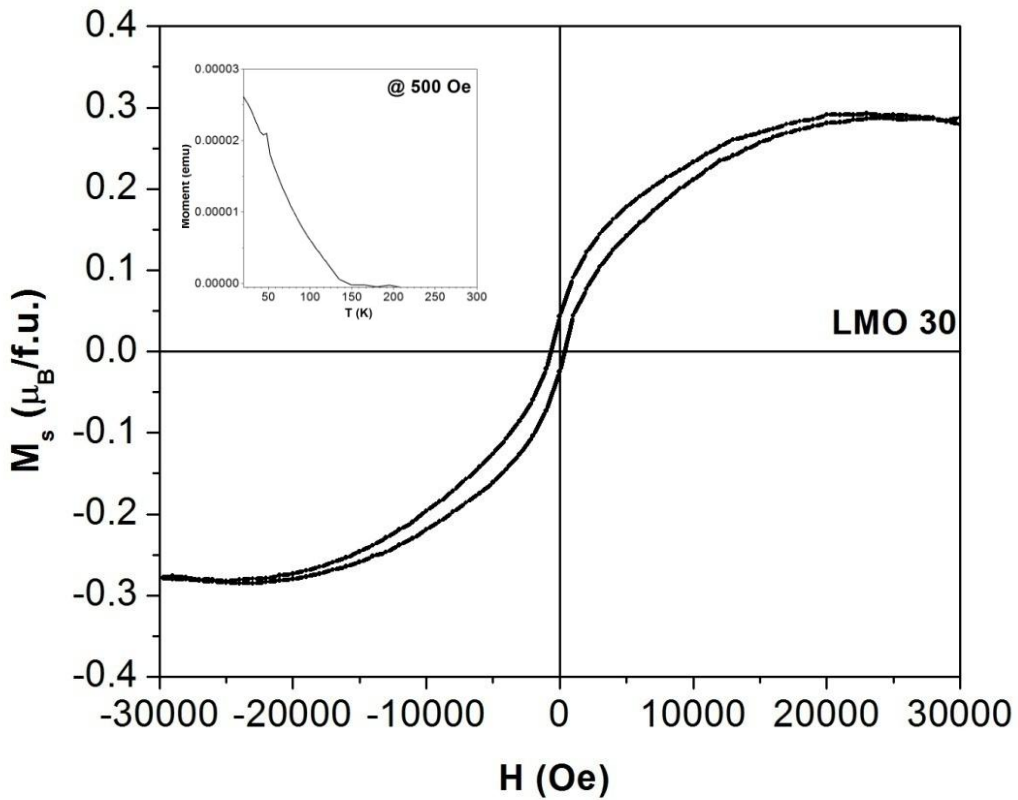

Figure 3.8.2.3. M-H loop for LMO 30 measured at 20K. The inset shows M-T curve measured at 500 Oe

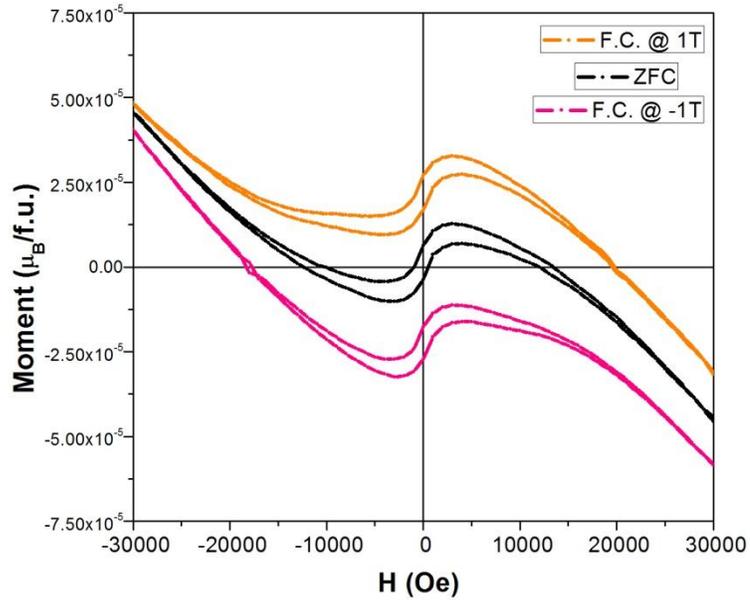

Figure 3.8.2.4. M-H loop with ZFC, FC(+1 Tesla) and FC(-1 Tesla)

## 3.9. LMO Thin Films Grown in Argon

LMO thin films grown in vacuum showed very good magnetic properties as discussed in previous sections. Vacuum growth confirms that LMO pellet doesn't need much reducing conditions as vacuum environment seems to be reducing for the LMO thin films. However, few LMO films were deposited in argon background to investigate and compare the magnetic properties with other growth environments.

### 3.9.1. Structural Properties of LMO Thin Films Grown in Argon

Higher laser fluence in this case leads to $La_2MnO_4$ phase. To investigate the effect of fluence two films were deposited with different pressure and at same fluence. The parameters which were kept constant for both the films are given in Table 3.9.1.1. Data from XRD and XRR are given in Table 3.9.1.2. with XRD θ-2θ scan as shown in Figure 3.9.1.1.

Table 3.9.1.1. Deposition parameters for different argon pressure

| Fluence | 1 J/cm$^2$ |
|---|---|
| Frequency | 2 |
| Temperature | 650 $^o$C |
| Target to Substrate Distance | 40 mm |
| Background | Argon |

Table 3.9.1.2. Thickness value of films deposited under same fluence in argon bacjground

| Film Number | Pressure (mTorr) | Thickness (nm) | Pulses |
|---|---|---|---|
| LMO 07 | 1 | 18.29 | 1000 |
| LMO 23 | 0.1 | 42.3 | 4000 |

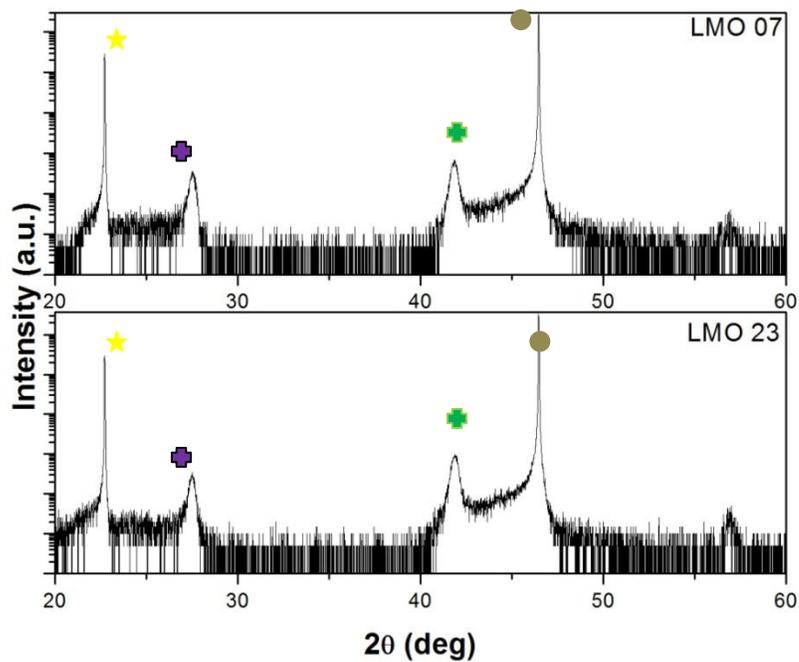

| ★ | STO (001) | ● | STO (002) |
|---|---|---|---|
| ✥ | La$_2$MnO$_4$ (001) | ✥ | La$_2$MnO$_4$ (002) |

Figure 3.9.1.1. XRD θ-2θ scan for LMO thin films grown in argon with different argon pressure and same fluence

As discussed earlier, higher laser fluence leads to $La_2MnO_4$ phase. So next step was to decrease the fluence at a constant pressure. The parameters which were kept constant for this scan, are given in Table 3.9.1.3. Data from XRD and XRR are given in Table 3.9.1.4. with XRD θ-2θ scan as shown in Figure 3.9.1.2.

Table 3.9.1.3. Deposition parameters for different fluence

| | |
|---|---|
| Pressure | 0.1 mTorr |
| Frequency | 2 |
| Pulses | 4000 |
| Temperature | 650 °C |
| Target to Substrate Distance | 40 mm |
| Background | Argon |

Argon background growth of LMO was not that easy to control with the available target. The target was stoichiometric with 'La' and 'Mn' in 3+ state. and argon atmosphere further reduces it and therefore films transform the phase to $La_2MnO_4$ [53-55].

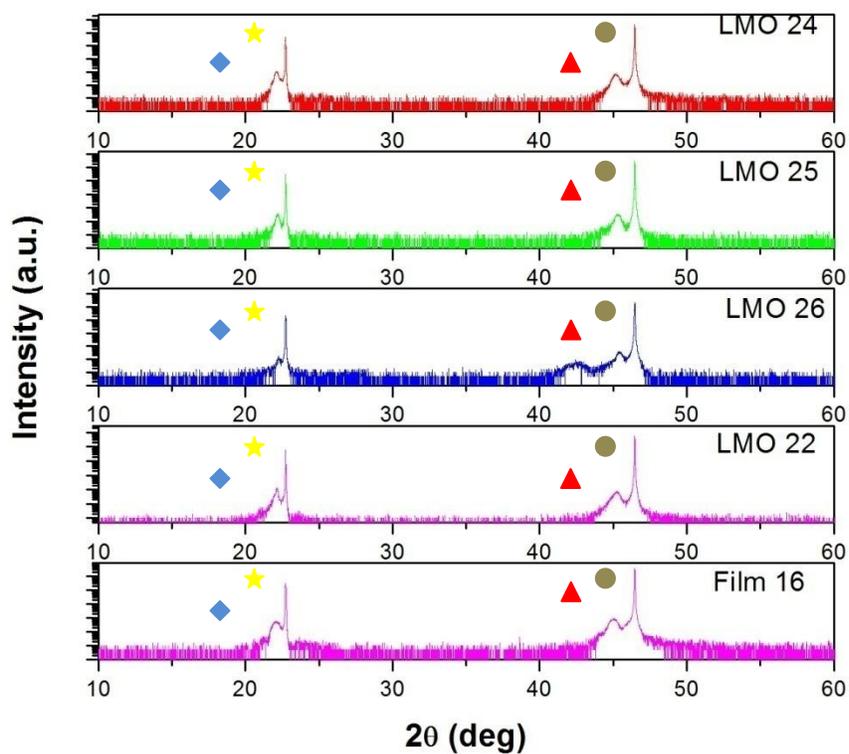

| ★ | STO (001) | ● | STO (002) |
|---|---|---|---|
| ◆ | LMO (001) | ▲ | LMO (002) |

Figure 3.9.1.2. XRD θ-2θ scan for LMO thin films grown in argon with different fluence at same pressure

Table 3.9.1.4. Thickness value of films deposited with different fluence at constant pressure

| Film Number | Fluence (J/cm$^2$) | Thickness (nm) |
|---|---|---|
| LMO 24 | 0.5 | 38.8 |
| LMO 25 | 0.6 | 49 |
| LMO 26 | 0.7 | 51 |
| LMO 22 | 0.75 | 56.5 |

### 3.9.2. Magnetic Properties of LMO Thin Films Grown in Argon

Magnetic properties were not that good in case of films grown in argon. The values of magnetic moments are given in Table 3.9.2.1 with M-H curves in figure 3.9.2.1.

Table 3.9.2.1. Magnetic moment for LMO thin films grown in argon with different fluence at same pressure

| Film Number | Fluence | Magnetic Moment | Pulses |
|---|---|---|---|
| LMO 24 | 0.5 | 0.52 $\mu_B$/f.u. | 4000 |
| LMO 25 | 0.6 | 0.41 $\mu_B$/f.u. | 4000 |
| LMO 26 | 0.7 | 0.6 $\mu_B$/f.u. | 4000 |
| LMO 22 | 0.75 | 0.5 $\mu_B$/f.u. | 4000 |
| LMO 16 | 0.75 | 0.5 $\mu_B$/f.u. | 1000 |

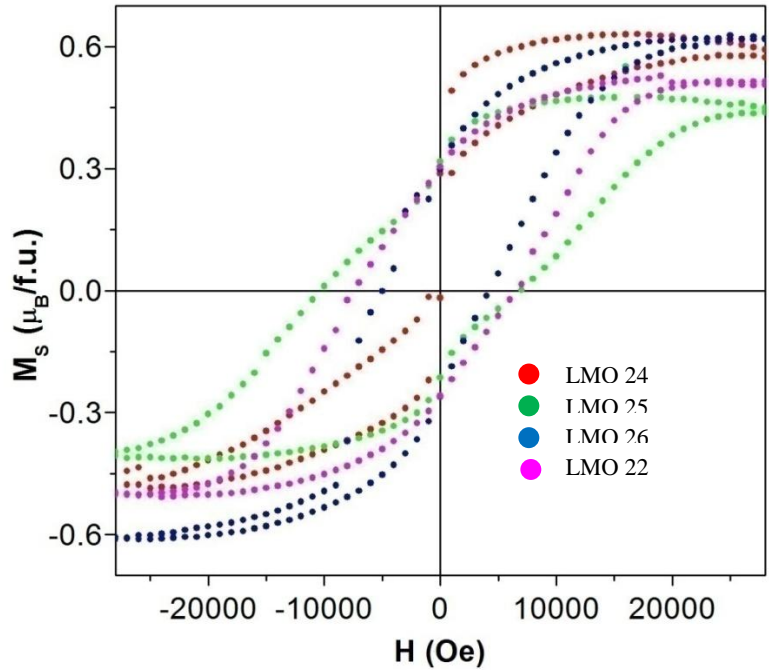

Figure 3.9.2.1. M-H loops measured at 20K for films grown in argon

## 3.10. Transport Properties of LMO Thin Films Grown in Oxygen and Vacuum

The transport properties as function of temperature is as shown in figure 3.10.1. The electrical resistivity decreases as the oxygen content decreases in the LMO thin film. The measurement setup which has been used to measure transport properties was very noisy below 135K. The sample grown in vacuum shows metal-to-insulator (MI) [56-58] transition at around 190 K and which decreases to 175 K when film grown in pure oxygen.

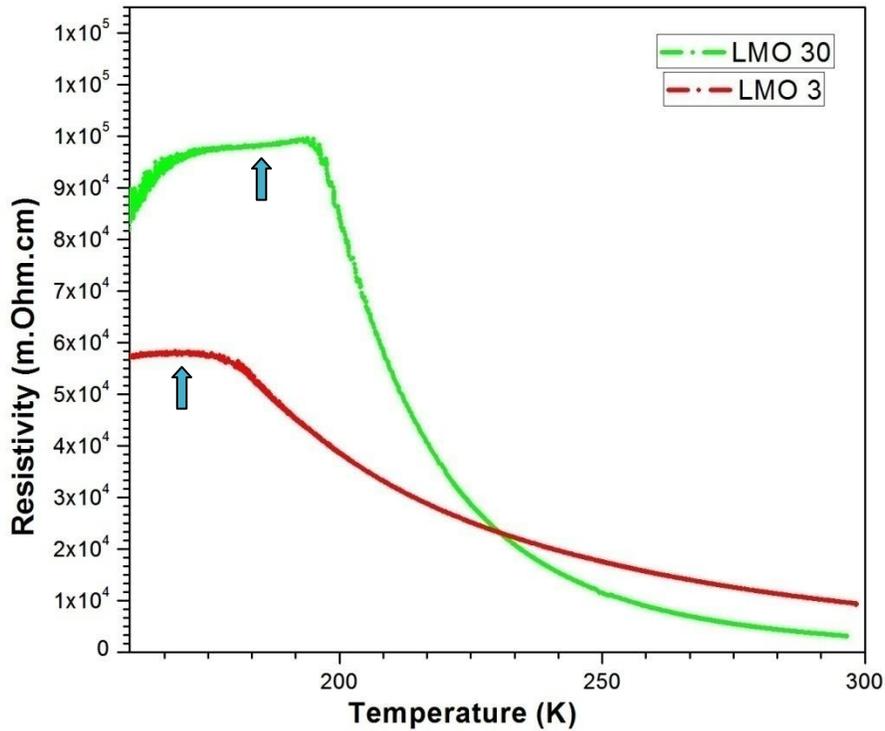

Figure 3.10.1. Temperature dependent resistivity measurement of LMO thin films grown in vacuum and oxygen

There exists relation between $Mn^{4+}$ concentration and the resistivity maxima, which occurs due to the metal-insulator transition [59]. Further investigation is needed on the dependence of resistivity measurements on the concentration of different ionic concentrations of Mn.

# CHAPTER 4
# CONCLUSION AND OUTLOOK

**Conclusion and Outlook**

This thesis work deals with the fabrication, characterization and optimization the growth conditions in order to minimize the ferromagnetic order in LMO thin films.

LMO target was prepared using conventional solid state reaction route. The XRD analysis of pellet confirms the LMO phase in addition to that small proportion of $La_{0.922}Mn_{1.013}O_3$ was also there. Cation stoichiometry was in accordance with the chemical formula. Density of pellet comes out to be 6 mg/cm$^3$ i.e, pellet was dense enough to be used for PLD process.

LMO thin films were deposited using pulsed laser deposition technique. Few films were deposited in oxygen background in order to check the electrical and magnetic properties and compare them with the available literature. Interestingly magnetic moment comes out to be smaller than the reported ones for LMO films grown in oxygen background. This was the first signature that observed stoichiometry was not that off from the desired stoichiometry.

LMO thin films were also deposited in vacuum. First a fluence scan was done and the results show, that a fluence of 0.5 J/cm$^2$ is the threshold above which phase transformation takes place. Fluence much higher than 0.5 J/cm$^2$ leads to either mixed phases or $La_2MnO_4$ phase. Afterward, thickness scan has been done for the fluence 0.5 J/cm$^2$. Results confirms the decrease in magnetic moment saturation as the thickness increases. Further reduction of fluence keeping all the other parameters constant leads to poor crystal quality. For all the films grown in vacuum, RHEED confirms layer by layer growth initially and then layer+Island growth. Phi scan confirms cubic-on-cubic growth of LMO thin films on STO(001) substrates. M-T curve shows the dependence of $T_c$ on thickness of the films. $T_c$ increases as the thickness increases.

Argon background leads to not so good magnetic properties. First fluence scan was done to find out the fluence threshold. The magnetic moments are higher than the films grown in vacuum. These conditions were too much reducing for the growth of LMO thin films with the available pellet.

Transport measurement results confirms the high resistivity of vacuum grown films as compared to films grown in oxygen. Films observed metal-to-insulator (MI) transition at different temperatures.

Al the films were reproducible and further optimization in vacuum and argon background is needed to further minimize the ferromagnetic order in thin films.

Further LMO thin films needs to be fabricated for the XPS studies to identify the effect of concentration of different ions on magnetic properties of films. Magnetic properties investigation of stoichiometric LMO thin films is also planned. Transport properties need to be investigated on lower temperatures.

# LIST OF PUBLICATIONS

1. **Control of stoichiometry in $LaMnO_3$/$La_2MnO_4$ thin films grown by pulsed laser deposition.**
Chouhan A S, Dasgupta S, Shabadi V, Radetinac A, Komissinskiy P, Thakur A D, Alff L. *Deutsche Physikalische Gesellschaft (DPG) e.V.,* **M19**, MA 19.60. (2015)

Official *URL* for 'Contribution' :
http://www.dpgverhandlungen.de/year/2015/conference/berlin/part/ma/session/19/contribution/60?lang=en